\RequirePackage{snapshot}
\documentclass[STIX1COL]{WileyNJD-v2}
\usepackage[T1]{fontenc}
\usepackage[utf8]{inputenc}
\usepackage{amsmath}
\usepackage{amsthm}
\usepackage{amssymb}
\usepackage{stmaryrd}
\usepackage{graphicx}
\usepackage{subscript}

\makeatletter

\articletype{Research Article}
\received{}
\revised{}
\accepted{} 
\jnlcitation{\cname{%
\author{E. J. Ching} and
\author{R. F. Johnson}}
(\cyear{2024}),
\ctitle{Conservative discontinuous Galerkin method for supercritical, real-fluid flows},
\cjournal{IJNMF},
\cvol{}} 

\providecommand{\tabularnewline}{\\}

\numberwithin{equation}{section}
\numberwithin{figure}{section}

\usepackage{tikz}
\usetikzlibrary{shapes,arrows}
\usepackage{float}

\@ifundefined{showcaptionsetup}{}{%
 \PassOptionsToPackage{caption=false}{subfig}}
\usepackage{subfig}
\makeatother

\begin{document}
\title{Conservative discontinuous Galerkin method for supercritical, real-fluid
flows}

\author[lcp]{Eric J. Ching}
\author[lcp]{Ryan F. Johnson}
\authormark{Ching and Johnson}
\address[lcp]{Laboratories for Computational Physics and Fluid Dynamics,  U.S. Naval Research Laboratory, 4555 Overlook Ave SW, Washington, DC 20375}
\corres{*Eric Ching, \email{eric.j.ching.civ@us.navy.mil}} 
\abstract{This paper presents a conservative discontinuous Galerkin method for
the simulation of supercritical and transcritical real-fluid flows
without phase separation. A well-known issue associated with the use
of fully conservative schemes is the generation of spurious pressure
oscillations at contact interfaces, which are exacerbated when a cubic
equation of state and thermodynamic relations appropriate for this
high-pressure flow regime are considered. To reduce these pressure
oscillations, which can otherwise lead to solver divergence in the
absence of additional dissipation, an $L^{2}$-projection of primitive
variables is performed in the evaluation of the flux. We apply the
discontinuous Galerkin formulation to a variety of test cases. The
first case is the advection of a sinusoidal density wave, which is
used to verify the convergence of the scheme. The next two involve
one- and two-dimensional advection of a nitrogen/n-dodecane thermal
bubble, in which the ability of the methodology to reduce pressure
oscillations and maintain solution stability is assessed. The final
test cases consist of two- and three-dimensional injection of an n-dodecane
jet into a nitrogen chamber.}

\maketitle
\global\long\def\middlebar{\,\middle|\,}%
\global\long\def\average#1{\left\{  \!\!\left\{  #1\right\}  \!\!\right\}  }%

\let\svthefootnote\thefootnote\let\thefootnote\relax\footnotetext{\\ \hspace*{110pt}DISTRIBUTION STATEMENT A. Approved for public release: distribution is unlimited.}\addtocounter{footnote}{-1}\let\thefootnote\svthefootnote

\section{Introduction}

\label{sec:Introduction}

Supercritical and transcritical flows are important in a variety of
engineering applications. For instance, in rocket engines, gas turbines,
and diesel engines, the fuel is often injected at high pressures to
improve engine efficiency and reduce emissions~\cite{Jaf22}. In
supercritical injection, the fuel is injected at a supercritical temperature.
On the other hand, in transcritical injection, the fuel is injected
at a subcritical temperature and then heated to supercritical temperatures
prior to combustion. Many of the relevant physical phenomena characterizing
these flows are not observed in subcritical systems. For example,
thermodynamic gradients are substantial, the liquid-gas phase boundary
begins to vanish (although phase separation may still occur in certain transcritical
flows), and surface tension may no longer play an important role.
Although various studies have been conducted to investigate supercritical
and transcritical flows, significant gaps in understanding of these
flows still exist.

Computational fluid dynamics (CFD) can help close these gaps, with
diffuse-interface methods commonly employed to simulate supercritical
and transcritical flows; however, certain numerical challenges hinder
the reliability of CFD for providing detailed insights into the relevant
physics. For instance, cubic or tabulated equations of state are often
employed, which increase complexity and can result in the loss of
certain desirable physical and mathematical properties associated
with the ideal-gas equation of state. Furthermore, the large density
gradients that exist between the gas-like and liquid-like phases easily
lead to numerical instabilities and potentially solver divergence.
One of the most prominent numerical challenges is the generation
of spurious pressure oscillations when fully conservative schemes
are employed. Commonly discussed in the context of multicomponent,
ideal-gas flows~\cite{Abg96}, this issue is caused by the loss of
mechanical equilibrium at contact interfaces due to variable thermodynamic
properties, which is further exacerbated by the increased nonlinearity
associated with supercritical and transcritical flows. In the absence
of appropriate numerical treatment, these spurious pressure oscillations
can grow rapidly and lead to solver divergence. A number of quasi-conservative
schemes have been developed to circumvent this problem. For instance,
a pressure evolution equation can be solved instead of the total-energy
equation to maintain pressure equilibrium~\cite{Kaw15}. The double-flux
scheme~\cite{Abg01}, which freezes the thermodynamics elementwise
and recasts the equation of state in a calorically-perfect form, is
another popular option. It was first developed for multicomponent,
ideal-gas flows and was later extended to transcritical flows~\cite{Ma17}.
However, the loss of energy conservation in these quasi-conservative
schemes is undesirable. To mitigate (but not eliminate) the conservation
error, Boyd and Jarrahbashi~\cite{Boy21} developed a hybrid scheme
that switches between the double-flux method and a fully conservative
method according to the jump in certain thermodynamic quantities.

Finite-difference and finite-volume schemes comprise the majority
of numerical techniques for simulating supercritical flows. The discontinuous
Galerkin (DG) method is a promising alternative due to its arbitrarily
high order of accuracy, compact stencil, geometric flexibility, compatibility
with local adaptivity, and amenability to modern computing architectures~\cite{Wan13}.
Apart from a handful of approaches in the literature~\cite{Hit20,Gry13},
DG-based formulations for transcritical and supercritical flows are
very limited. Given the advantages offered by DG schemes, the main
objectives of this work are to present a fully conservative DG formulation~\cite{Joh20_2,Chi23_short}
for supercritical and transcritical flows (without phase separation)
and assess its ability to accurately and robustly simulate flows in
this regime. This formulation was originally developed to simulate
multicomponent, chemically reacting flows involving mixtures of thermally
perfect gases~\cite{Joh20_2}. It was found that using a colocated
scheme, pressure equilibrium was successfully maintained (in an approximate
sense) in the one-dimensional advection of a hydrogen/oxygen thermal
bubble. Although pressure oscillations were present, they remained
small throughout the simulation. In contrast, with standard overintegration,
pressure oscillations grew rapidly and caused solver divergence. To
maintain solution stability while employing overintegration, the authors
introduced an overintegration strategy in which the pressure is projected
onto the finite element test space via interpolation~\cite{Joh20_2}.
Bando~\cite{Ban23} further assessed this overintegration strategy
and compared it to a modified strategy, inspired by a similar approach
by Franchina et al.~\cite{Fra16}, in which the pressure is projected
instead via $L^{2}$-projection. Using the same hydrogen/oxygen thermal-bubble
configuration as a test bed, it was concluded that although both approaches
maintained approximate pressure equilibrium, the interpolation-based
approach is preferred since it is simpler and yielded smaller pressure
oscillations. However, it was recently found that in a more challenging
test case involving the one-dimensional advection of a high-pressure,
ntirogen/n-dodecane thermal bubble (which is more representative of
realistic transcritical/supercritical flows even though the thermally
perfect gas model was employed), the $L^{2}$-projection-based overintegration
strategy significantly improved solution stability~\cite{Chi23_short}.
 Here, we evaluate whether this overintegration strategy can similar
maintain solution stability when real-fluid effects in multiple dimensions
are considered.

The remainder of this paper is organized as follows. Sections~\ref{sec:Governing-equations}
and~\ref{sec:DG-discretization} summarize the governing equations
and DG formulation, respectively. Results for a variety of test cases,
including multidimensional advection of a nitrogen/n-dodecane thermal
bubble and injection of an n-dodecane jet into a nitrogen chamber,
are given in the following section. We close the paper with concluding
remarks.

\section{Governing equations}

\label{sec:Governing-equations}

The governing equations for conservation of species concentrations,
momentum, and total energy are written as
\begin{equation}
\frac{\partial y}{\partial t}+\nabla\cdot\mathcal{F}\left(y\right)=0,\label{eq:conservation-law-strong-form}
\end{equation}
where $y$ is the vector of $m$ state variables, $t$ is time, and
$\mathcal{F}$ is the convective flux. The physical coordinates are
denoted by $x=(x_{1},\ldots,x_{d})$, where $d$ is the number of
spatial dimensions. The state vector is expanded as
\begin{equation}
y=\left(\rho v_{1},\ldots,\rho v_{d},\rho e_{t},C_{1},\ldots,C_{n_{s}}\right)^{T},\label{eq:reacting-navier-stokes-state}
\end{equation}
where $\rho$ is the density, $v=\left(v_{1},\ldots,v_{d}\right)$
is the velocity vector, $e_{t}$ is the specific total energy, $C=\left(C_{1},\ldots,C_{n_{s}}\right)$
is the vector of molar concentrations, and $n_{s}$ is the number
of species. The density can be computed as
\[
\rho=\sum_{i=1}^{n_{s}}\rho_{i}=\sum_{i=1}^{n_{s}}W_{i}C_{i},
\]
where $\rho_{i}$ and $W_{i}$ are the partial density and molecular
weight, respectively, of the $i$th species. The mole and mass fractions
of the $i$th species are given by
\[
X_{i}=\frac{C_{i}}{\sum_{i=1}^{n_{s}}C_{i}},\quad Y_{i}=\frac{\rho_{i}}{\rho}.
\]
The specific total energy is expanded as 
\[
e_{t}=u+\frac{1}{2}\sum_{k=1}^{d}v_{k}v_{k},
\]
where $u$ is the mixture-averaged specific internal energy, computed
as the mass-weighted sum of the specific internal energies of each
species, i.e.,

\[
u=\sum_{i=1}^{n_{s}}Y_{i}u_{i}.
\]
 The $k$th spatial component of the convective flux is given by
\begin{equation}
\mathcal{F}_{k}^ {}\left(y\right)=\left(\rho v_{k}v_{1}+P\delta_{k1},\ldots,\rho v_{k}v_{d}+P\delta_{kd},v_{k}\left(\rho e_{t}+P\right),v_{k}C_{1},\ldots,v_{k}C_{n_{s}}\right)^{T},\label{eq:reacting-navier-stokes-spatial-convective-flux-component}
\end{equation}
where $P$ is the pressure, which is computed using the Peng-Robinson
cubic equation of state~\cite{Pen76}:
\begin{equation}
P=\frac{\widehat{R}T}{\widehat{v}-b}-\frac{a\alpha}{\widehat{v}^{2}+2b\widehat{v}-b^{2}}.\label{eq:peng-robinson}
\end{equation}
$T$ is the temperature, $\widehat{v}=1/\sum_{i=1}^{n_{s}}C_{i}$
is the molar volume, $a\alpha$ and $b$ are the attractive and repulsive
parameters, respectively, and $\widehat{R}$ is the universal gas
constant. To further account for real-fluid effects, thermodynamic
quantities are computed by augmenting the NASA-polynomial-based ideal-gas
values~\cite{Mcb93,Mcb02} with departure functions, which, in the
case of mixtures, are evaluated using the extended corresponding states
principle and pure-fluid assumption~\cite{Ely81,Ely83} and the recommended
mixing rules by Harstad et al.~\cite{Har97}. More information on
the exact forms of the departure functions for a generic cubic equation
of state can be found in Appendix~\ref{sec:thermo-relations}. As
in~\cite{Boy21}, viscous effects are ignored here since they are
not the cause of spurious pressure oscillations. Surface tension is
also neglected, which is typical for supercritical/transcritical flow
models~\cite{Ma17,Boy21,Ma18}. Table~\ref{tab:species-properties}
lists the critical properties of the species considered in Section~\ref{sec:results}. 

\begin{table}[h]
\begin{centering}
\caption{Relevant properties of the species considered in this paper. $W$
is the molecular mass, $\left(\cdot\right)_{c}$ denotes critical
property, and $\omega$ is the acentric factor.\label{tab:species-properties}}
\par\end{centering}
\centering{}%
\begin{tabular}{cccccc}
\hline 
Species & $W$ (kg/kmol) & $T_{c}$ (K) & $P_{c}$ (MPa) & $\rho_{c}$ (kg/m\textsuperscript{3}) & $\omega$\tabularnewline
\hline 
N\textsubscript{2} & 28.0 & 126.2 & 3.40 & 313.3 & 0.0372\tabularnewline
\emph{n}-C\textsubscript{12}H\textsubscript{26} & 170.3 & 658.1 & 1.82 & 226.5 & 0.574\tabularnewline
\end{tabular}
\end{table}

A well-known issue with the Peng-Robinson equation of state is that
it returns negative pressure and complex sound speed under certain
conditions. Given the flows targeted in this study (i.e., no phase
separation), such conditions are obtained typically as a result of
numerical instabilities, in which case we limit the pressure and speed
of sound returned by the equation of state to minimum values of 10
Pa and 1 m/s, respectively. More sophisticated modifications can also
be employed~\cite{Boy21,Ma19_2}.

\section{Discontinuous Galerkin discretization}

\label{sec:DG-discretization}

Let $\Omega$ denote the computational domain partitioned by $\mathcal{T}$,
which consists of cells $\kappa$ with boundaries $\partial\kappa$.
Let $\mathcal{E}$ denote the set of interfaces $\epsilon$, consisting
of the interior interfaces,
\[
\epsilon_{\mathcal{I}}\in\mathcal{E_{I}}=\left\{ \epsilon_{\mathcal{I}}\in\mathcal{E}\middlebar\epsilon_{\mathcal{I}}\cap\partial\Omega=\emptyset\right\} ,
\]
and boundary interfaces, 
\[
\epsilon_{\partial}\in\mathcal{E}_{\partial}=\left\{ \epsilon_{\partial}\in\mathcal{E}\middlebar\epsilon_{\partial}\subset\partial\Omega\right\} .
\]
At interior interfaces, there exist $\kappa^{+}$ and $\kappa^{-}$
such that $\epsilon_{\mathcal{I}}=\partial\kappa^{+}\cap\partial\kappa^{-}$.
$n^{+}$ and $n^{-}$ denote the outward facing normals of $\kappa^{+}$
and $\kappa^{-}$, respectively. Let $V_{h}^{p}$ denote the space
of test and basis functions
\begin{eqnarray}
V_{h}^{p} & = & \left\{ \mathfrak{v}\in\left[L^{2}\left(\Omega\right)\right]^{m}\middlebar\forall\kappa\in\mathcal{T},\left.\mathfrak{v}\right|_{\kappa}\in\left[\mathcal{P}_{p}(\kappa)\right]^{m}\right\} ,\label{eq:discrete-subspace}
\end{eqnarray}
where $\mathcal{P}_{p}(\kappa)$ is a space of polynomial functions
of degree no greater than $p$ in $\kappa$.

The semi-discrete form of Equation~(\ref{eq:conservation-law-strong-form})
is as follows: find $y\in V_{h}^{p}$ such that
\begin{gather}
\sum_{\kappa\in\mathcal{T}}\left(\frac{\partial y}{\partial t},\mathfrak{v}\right)_{\kappa}-\sum_{\kappa\in\mathcal{T}}\left(\mathcal{F}\left(y\right),\nabla\mathfrak{v}\right)_{\kappa}+\sum_{\epsilon\in\mathcal{E}}\left(\mathcal{F}^{\dagger}\left(y^{+},y^{-},n\right),\left\llbracket \mathfrak{v}\right\rrbracket \right)_{\mathcal{E}}=0\qquad\forall\:\mathfrak{v}\in V_{h}^{p},\label{eq:semi-discrete-form}
\end{gather}
where $\left(\cdot,\cdot\right)$ denotes the inner product, $\mathcal{F}^{\dagger}\left(y^{+},y^{-},n\right)$
is the numerical flux (taken to be the HLLC flux function~\cite{Tor13}),
and $\left\llbracket \cdot\right\rrbracket $ is the jump operator,
given by $\left\llbracket \mathfrak{v}\right\rrbracket =\mathfrak{v}^{+}-\mathfrak{v}^{-}$
at interior interfaces and $\left\llbracket \mathfrak{v}\right\rrbracket =\mathfrak{v}^{+}$
at boundary interfaces. Throughout this work, a nodal basis is employed,
such that the element-local polynomial approximation of the solution
is expanded as
\begin{equation}
y_{\kappa}=\sum_{j=1}^{n_{b}}y_{\kappa}(x_{j})\phi_{j},\label{eq:solution-approximation}
\end{equation}
where $n_{b}$ is the number of basis functions, $\left\{ \phi_{1},\ldots,\phi_{n_{b}}\right\} $
is the set of basis functions, and $\left\{ x_{1},\ldots,x_{n_{b}}\right\} $
is the set of node coordinates. 

\subsection{Integration\label{subsec:Integration}}

The integrals in Equation~(\ref{eq:semi-discrete-form}) are computed
using a quadrature-free approach~\cite{Atk96,Atk98}. In the evaluation
of the second and third integrals in Equation~(\ref{eq:semi-discrete-form}),
the nonlinear convective flux is approximated as either
\begin{equation}
\mathcal{F_{\kappa}}\approx\sum_{k=1}^{n_{b}}\mathcal{F}\left(y_{\kappa}\left(x_{k}\right)\right)\phi_{k},\label{eq:flux-projection-colocated}
\end{equation}
in the case of colocated integration, or
\begin{equation}
\mathcal{F_{\kappa}}\approx\sum_{k=1}^{n_{c}}\mathcal{F}\left(\left.\text{\ensuremath{\mathcal{P}}}\left(z\left(y_{\kappa}\right)\right)\right|_{x_{k}}\right)\varphi_{k},\label{eq:flux-projection-overintegration}
\end{equation}
in the case of overintegration, where $n_{c}>n_{b}$, $\left\{ \varphi_{1},\ldots,\varphi_{n_{c}}\right\} $
is the corresponding set of basis functions, $\mathcal{P}$ is a projection
operator, and $z\left(y\right):\mathbb{R}^{m}\rightarrow\mathbb{R}^{m}$
is a vector of intermediate state variables. In previous work~\cite{Chi23_short},
a number of options for $\mathcal{P}$ and $z$ were investigated
in the context of mixtures of thermally perfect gases. Based on those
results, we choose $\mathcal{P}=\Pi$, which is the $L^{2}$-projection
operator onto $V_{h}^{p}$, and $z=\left(v_{1},\ldots,v_{d},P,C_{1},\ldots,C_{n_{s}}\right)^{T}$.
Other choices of $\mathcal{P}$, specifically the identity function
(which corresponds to standard flux evaluation, i.e., $\sum_{k=1}^{n_{c}}\mathcal{F}\left(y_{\kappa}\left(x_{k}\right)\right)\varphi_{k}$)
and interpolatory projection onto $V_{h}^{p}$, were not sufficiently
robust in~\cite{Chi23_short}. Note that in~\cite{Chi23_short},
$z=\left(v_{1},\ldots,v_{d},P,T,Y_{1},\ldots,Y_{n_{s}-1}\right)^{T}$
yielded the smallest deviations from pressure equilibrium; however,
we find that this choice of $z$ excessively modifies species concentrations,
especially in Sections~\ref{subsec:Two-dimensional-n-dodecane-jet}
and~\ref{subsec:Three-dimensional-n-dodecane-jet}.

\subsection{Stabilization}

Artificial viscosity is employed to stabilize the solution at discontinuous
interfaces in Section~\ref{subsec:Two-dimensional-n-dodecane-jet}.
Note that we are primarily interested in contact interfaces in this
work. Transcritical/supercritical flows with shocks are outside the
scope of this study. The following dissipation term is added to the
RHS of Equation~(\ref{eq:semi-discrete-form})~\cite{Har13}:
\begin{equation}
\sum_{\kappa\in\mathcal{T}}\left(\mathcal{F}^{\mathrm{AV}}\left(y,\nabla y\right),\nabla\mathfrak{v}\right)_{\kappa},\label{eq:artificial-viscosity-integral}
\end{equation}
where
\[
\mathcal{F}^{\mathrm{AV}}(y,\nabla y)=\nu_{\mathrm{AV}}\left(y\right)\nabla y,
\]
with $\nu_{\mathrm{AV}}\geq0$ denoting the artificial viscosity,
which is computed as~\cite{Joh20_2}
\[
\nu_{\mathrm{AV}}=\left(C_{\mathrm{AV}}+S_{\mathrm{AV}}\right)\left(\frac{h^{2}}{p+1}\left|\frac{\partial T}{\partial y}\cdot\frac{\mathcal{R}\left(y,\nabla y\right)}{T}\right|\right).
\]
$C_{\mathrm{AV}}$ is a user-defined coefficient, $S_{\mathrm{AV}}$
is a sensor based on intra-element variations~\cite{Chi19}, and
$\mathcal{R}\left(y,\nabla y\right)$ is the strong form of the residual~(\ref{eq:conservation-law-strong-form}).
This type of artificial viscosity was found to effectively suppress
spurious oscillations in the vicinity of flow-field discontinuities
in multicomponent reacting flows~\cite{Joh20_2,Chi22,Chi22_2}. Note,
however, that the artificial-viscosity formulation described here
is not the focus of this paper. Investigating other dissipative stabilization
techniques, such as filtering and limiters, that may more effectively
dampen nonlinear instabilities will be the subject of future work. 

Finally, we employ a simple linear-scaling limiter~\cite{Zha10,Chi22,Chi22_2}
to help ensure well-behaved species concentrations, densities, and
temperatures. It should be noted, however, that the formulation is
not mathematically guaranteed to be positivity-preserving. Furthermore,
this limiter by itself does not suppress small-scale instabilities. 

\section{Results}

\label{sec:results}

We consider four test cases. In the first one, which involves a sinusoidal
density wave, we verify optimal convergence of the formulation. The
second case is the one-dimensional advection of a nitrogen/n-dodecane
thermal bubble, where we assess the ability of the methodology to
maintain pressure equilibrium and solution stability. The third test
is a two-dimensional version of the previous one. Curved elements
are also considered. No additional dissipation is applied to the computations
of the aforementioned test cases. The final two configurations involve
the injection of an n-dodecane jet into a nitrogen chamber in two
and three dimensions. All solutions are initialized using interpolation
and then integrated in time using the third-order strong-stability-preserving
Runge-Kutta scheme~\cite{Got01}. All simulations are performed using
a modified version of the JENRE\textregistered~Multiphysics Framework~\cite{Joh20_2}
that incorporates the extensions described in this work.

\subsection{Sinusoidal density wave\label{subsec:density-wave}}

\begin{figure}[h]
\subfloat[\label{fig:convergence_colocated}Colocated integration.]{\includegraphics[width=0.48\columnwidth]{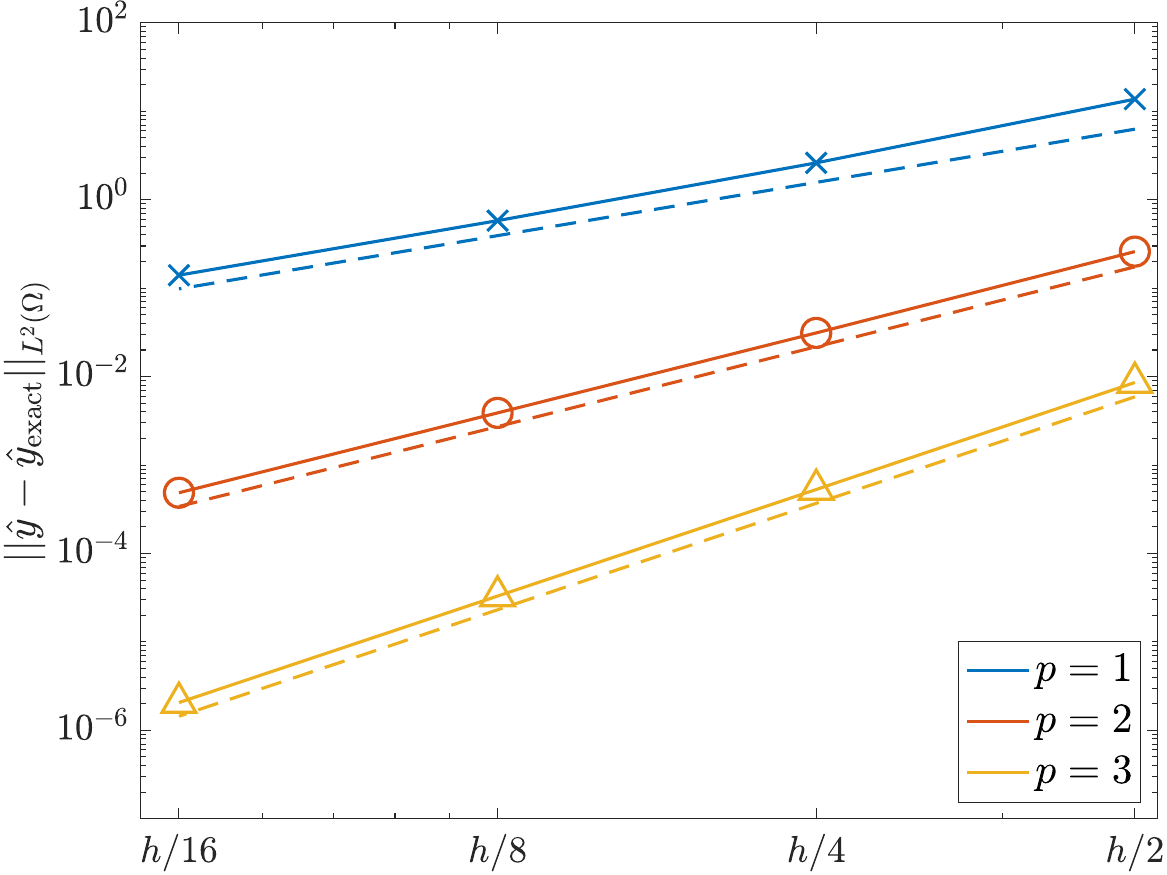}}\hfill{}\subfloat[\label{fig:convergence_overintegrated}Overintegration.]{\includegraphics[width=0.48\columnwidth]{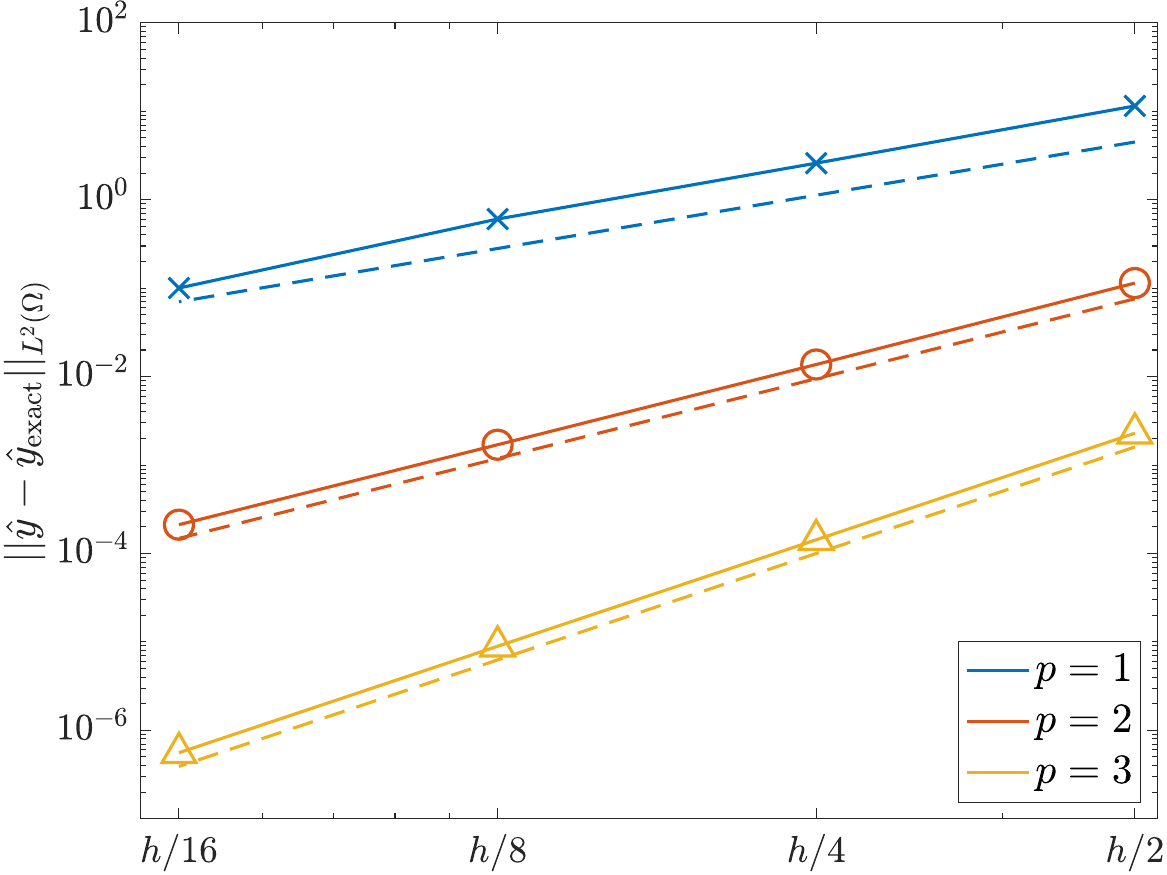}}

\caption{\label{fig:convergence}Convergence under grid refinement, with $h=0.04\:\mathrm{m}$,
for the one-dimensional sinusoidal density wave test. Figure~\ref{fig:convergence_colocated}
corresponds to colocated integration, and Figure~\ref{fig:convergence_overintegrated}
corresponds to the overintegration strategy discussed in Section~\ref{subsec:Integration}.
The $L^{2}$ error of the normalized state with respect to the exact
solution is computed. The dashed lines represent the theroretical
convergence rates.}
\end{figure}

As in~\cite{Ma17}, we use this smooth flow problem to assess the
grid convergence of the DG formulation (without artificial viscosity).
A one-dimensional periodic domain $\Omega\in\left[0,1\right]\text{ m}$
is initialized as follows:

\begin{eqnarray*}
v_{1} & = & 100\textrm{ m/s},\\
Y_{N_{2}} & = & 1,\\
\rho & = & \frac{\rho_{\min}+\rho_{\max}}{2}-\frac{\rho_{\max}-\rho_{\min}}{2}\sin\left(2\pi x\right)\textrm{ kg/m}^{3},\\
P & = & 5\textrm{ MPa},
\end{eqnarray*}
where $\rho_{\min}=56.9~\mathrm{kg/m^{3}}$ and $\rho_{\min}=794~\mathrm{kg/m^{3}}$,
which correspond to $T_{\max}=300~\mathrm{K}$ and $T_{\min}=100~\mathrm{K}$,
respectively. Four element sizes are considered: $h$, $h/2$, $h/4$,
and $h/8$, where $h=0.04\:\mathrm{m}$. We use a CFL number of 0.1
(based on the order-dependent linear-stability constraint) to minimize
temporal error. The $L^{2}$ error at $t=0.01\:\mathrm{s}$, which
corresponds to one period, is calculated in terms of the following
normalized state variables:
\[
\widehat{\rho v}_{k}=\frac{1}{\sqrt{\rho_{r}P_{r}}}\rho v_{k},\quad\widehat{\rho e}_{t}=\frac{1}{P_{r}}\rho e_{t},\quad\widehat{C}_{i}=\frac{R^{0}T_{r}}{P_{r}}C_{i},
\]
where $T_{r}=1000\,\mathrm{K}$, $\rho_{r}=1\,\mathrm{kg\cdot}\mathrm{m}^{-3}$,
and $P_{r}=101325\,\mathrm{Pa}$ are reference values. The exact solution
is simply the initial condition. We assess convergence with both colocated
integration based on Gauss-Lobatto nodal sets and the overintegration
strategy discussed in Section~\ref{subsec:Integration}. Figure~\ref{fig:convergence}
presents results for $p=1$ to $p=3$. The dashed lines represent
the theoretical convergence rates. The errors are slightly lower with
the overintegration~(\ref{eq:flux-projection-overintegration}).
Optimal order of accuracy is observed. 

\subsection{One-dimensional advection of nitrogen/n-dodecane thermal bubble\label{subsec:thermal_bubble}}

This section presents results for the one-dimensional advection of
a nitrogen/n-dodecane thermal bubble~\cite{Ma17,Boy21}. In previous
work~\cite{Chi23_short}, we computed this test case with the thermally
perfect gas model.  We are interested in how the proposed formulation
performs here given the additional nonlinearities associated with
the cubic equation of state and more complicated thermodynamic relations.
The initial conditions are given by
\begin{eqnarray}
Y_{n\text{-}\mathrm{C_{12}H_{26}}} & = & \frac{1}{2}\left[1-\tanh\left(25|x|-5\right)\right],\nonumber \\
Y_{\mathrm{N_{2}}} & = & 1-Y_{n\text{-}\mathrm{C_{12}H_{26}}},\nonumber \\
T & = & \frac{T_{\min}+T_{\max}}{2}+\frac{T_{\max}-T_{\min}}{2}\tanh\left(25|x|-5\right)\textrm{ K},\label{eq:thermal-bubble}\\
P & = & 6\textrm{ MPa},\nonumber \\
v_{1} & = & 1\text{ m/s}\nonumber 
\end{eqnarray}
where $T_{\min}=363\text{ K}$ and $T_{\max}=900\text{ K}$. The
periodic domain is $\Omega=\left[-0.5,0.5\right]\text{ m}$. The HLLC~\cite{Tor13}
flux function is employed. All solutions are advanced in time for
ten periods with $\mathrm{CFL}=0.8$. Two separate grids with cell
sizes of $h=0.01~\mathrm{m}$ and $h=0.005~\mathrm{m}$ are used.
No artificial viscosity is employed.

\begin{figure}[h]
\subfloat[\label{fig:thermal_bubble_dodecane_P_p1}Pressure, $p=1$.]{\includegraphics[width=0.48\columnwidth]{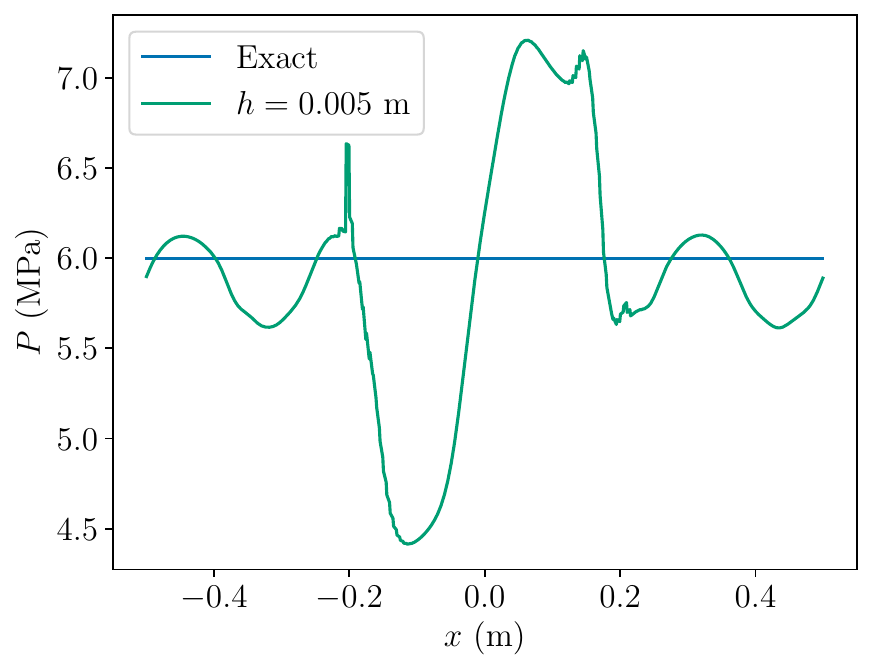}}\hfill{}\subfloat[\label{fig:thermal_bubble_dodecane_T_p1}Temperature, $p=1$.]{\includegraphics[width=0.48\columnwidth]{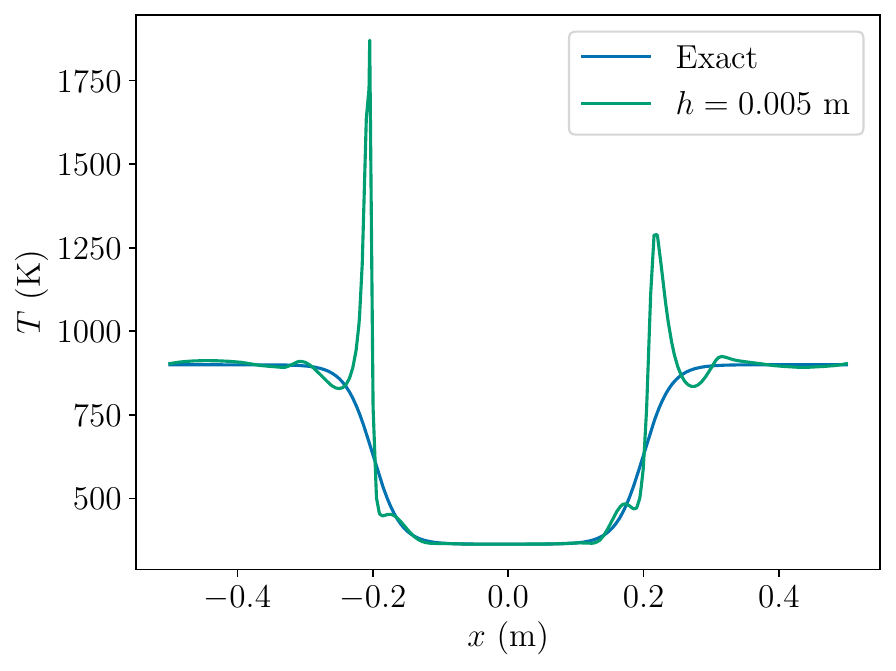}}

\hfill{}\subfloat[\label{fig:thermal_bubble_dodecane_P_p2}Pressure, $p=2$.]{\includegraphics[width=0.48\columnwidth]{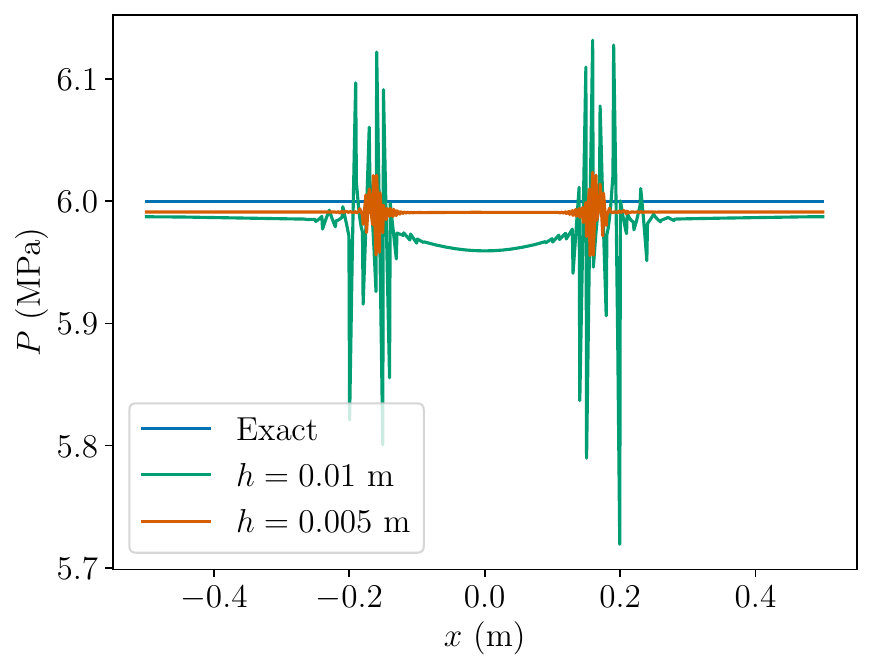}}\hfill{}\subfloat[\label{fig:thermal_bubble_dodecane_T_p2}Temperature, $p=2$.]{\includegraphics[width=0.48\columnwidth]{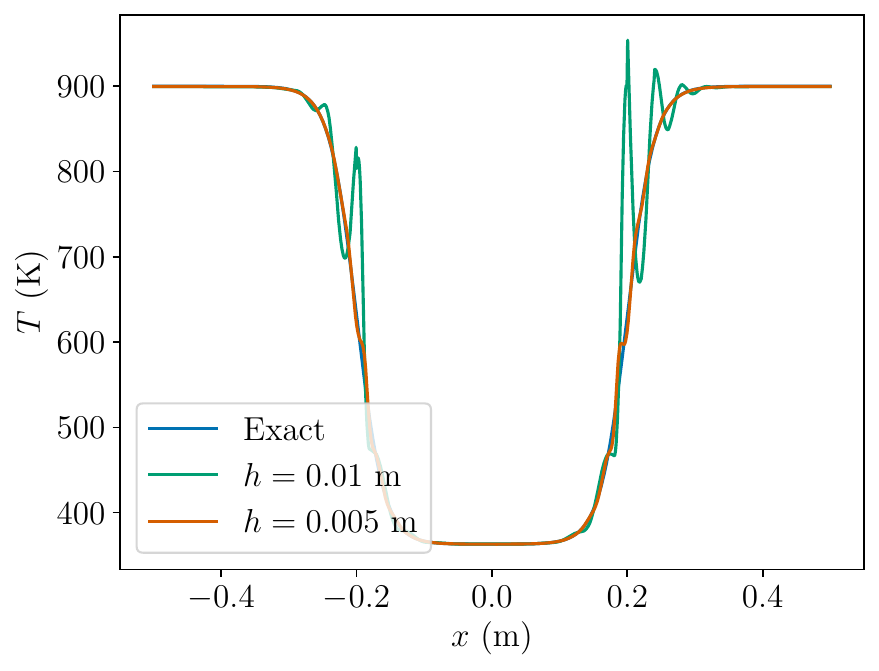}}\hfill{}\subfloat[\label{fig:thermal_bubble_dodecane_P_p3}Pressure, $p=3$.]{\includegraphics[width=0.48\columnwidth]{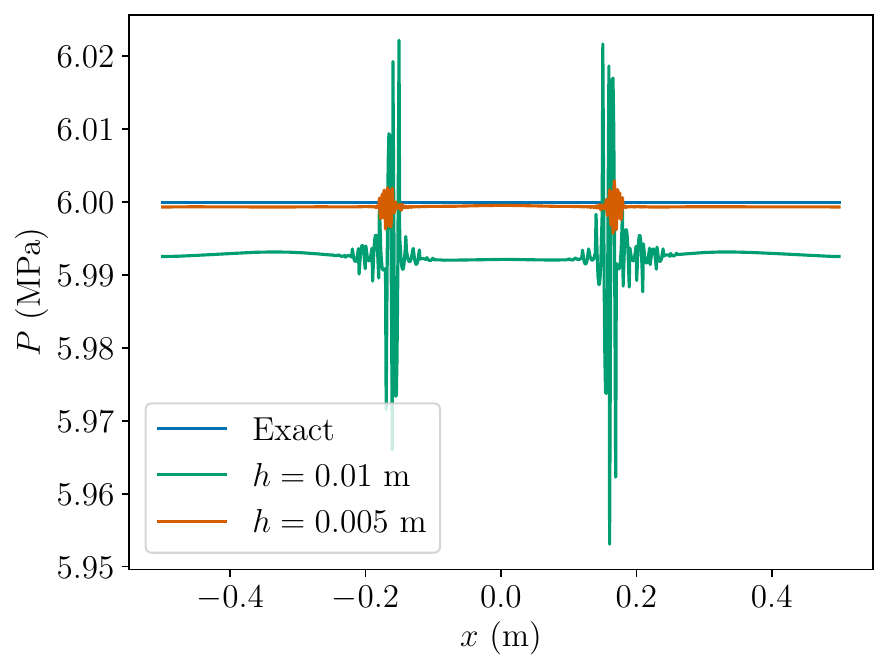}}\hfill{}\subfloat[\label{fig:thermal_bubble_dodecane_T_p3}Temperature, $p=3$.]{\includegraphics[width=0.48\columnwidth]{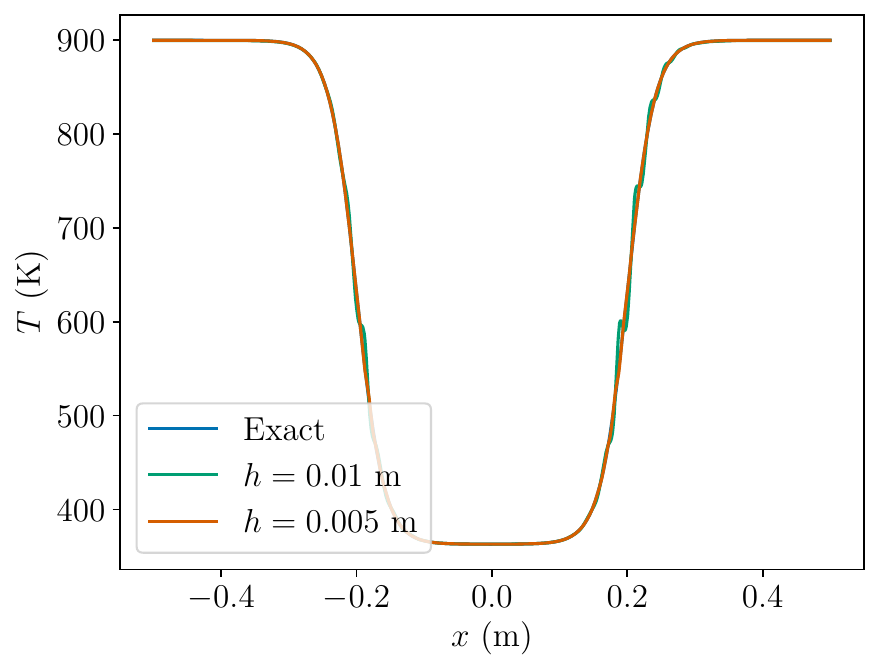}}

\caption{\label{fig:thermal_bubble_dodecane}$p=1$, $p=2$, and $p=3$ solutions
to the one-dimensional advection of a nitrogen/n-dodecane thermal
bubble after ten periods. The overintegration~(\ref{eq:flux-projection-overintegration})
is employed. Pressure and temperature profiles obtained with cell
sizes of $h=0.02~\mathrm{m}$ and $h=0.01~\mathrm{m}$. However, the
$p=1,h=0.02~\mathrm{m}$ solution diverges and is therefore not shown.}
\end{figure}

Figure~\ref{fig:thermal_bubble_dodecane} presents $p=1$, $p=2$,
and $p=3$ solutions obtained with the overintegration~(\ref{eq:flux-projection-overintegration}).
All solutions remain stable except for the $p=1$ solution on the
coarser grid. With standard overintegration, all solutions become
unstable. $p=0$ solutions on both meshes diverge as well.   This
illustrates the increased difficulty of maintaining solution robustness
when real-fluid effects are considered. Despite noticeable deviations
from pressure equilibrium in the coarser cases, they do not grow unbounded,
as they would if standard integration were employed. As the polynomial
order and/or number of cells increases, pressure equilibrium is better
maintained and temperature is more accurately predicted. Note that
despite the same number of degrees of freedom, the $p=3$ solution
on the coarser grid is more accurate than the $p=1$ solution on the
finer grid.

\subsection{Two-dimensional advection of nitrogen/n-dodecane thermal bubble}

\begin{figure}[h]
\subfloat[\label{fig:thermal_bubble_2d_initial_density}Initial density field.]{\includegraphics[width=0.48\columnwidth]{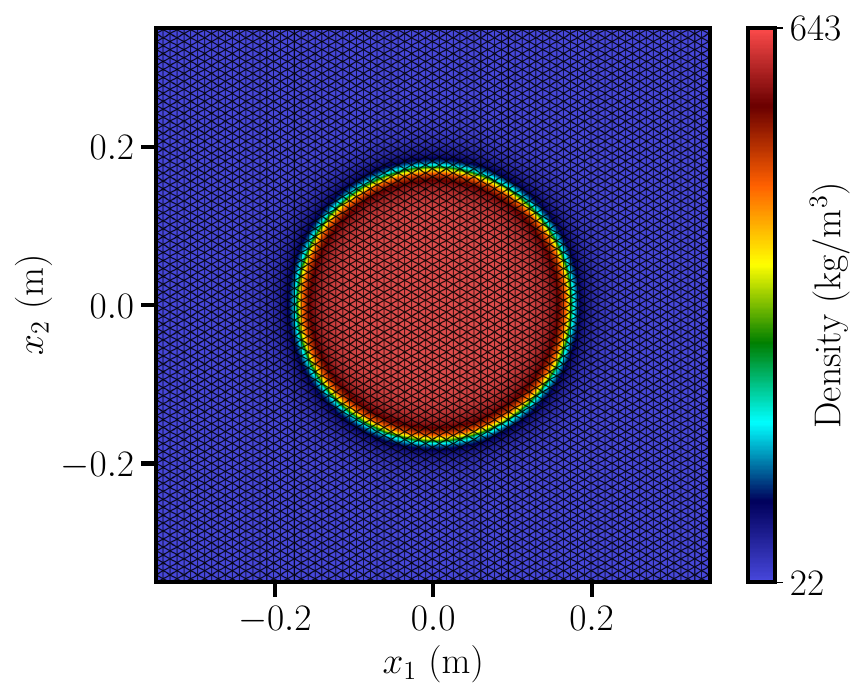}}\hfill{}\subfloat[\label{fig:thermal_bubble_2d_initial_temperature}Initial temperature
field.]{\includegraphics[width=0.48\columnwidth]{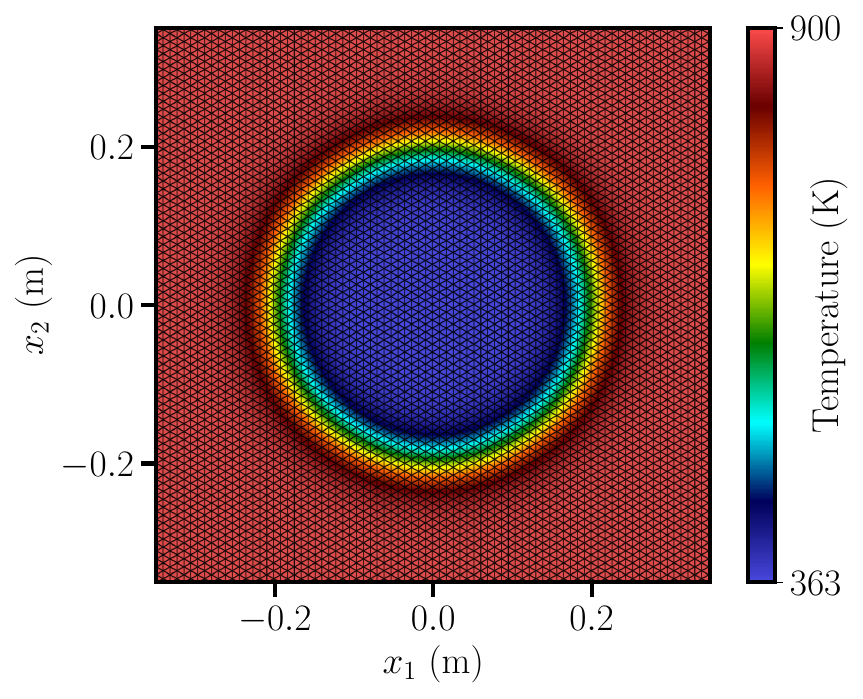}}

\caption{\label{fig:thermal_bubble_2d_initial} Initial $p=3$ density and
pressure fields, superimposed by the grid, for the two-dimensional
advection of a nitrogen/n-dodecane thermal bubble.}
\end{figure}

This flow configuration is a two-dimensional version of the previous
test case, with initial conditions written as
\begin{eqnarray}
Y_{n\text{-}\mathrm{C_{12}H_{26}}} & = & \frac{1}{2}\left[1-\tanh\left(25\sqrt{x_{1}^{2}+x_{2}^{2}}-5\right)\right],\nonumber \\
Y_{\mathrm{N_{2}}} & = & 1-Y_{n\text{-}\mathrm{C_{12}H_{26}}},\nonumber \\
T & = & \frac{T_{\min}+T_{\max}}{2}+\frac{T_{\max}-T_{\min}}{2}\tanh\left(25\sqrt{x_{1}^{2}+x_{2}^{2}}-5\right)\textrm{ K},\label{eq:thermal-bubble-2d}\\
P & = & 6\textrm{ MPa},\nonumber \\
\left(v_{1},v_{2}\right) & = & \left(600,0\right)\textrm{ m/s},\nonumber 
\end{eqnarray}
Note that the velocity is higher than in the previous test. In previous
work~\cite{Chi23_short}, we investigated this high-velocity case
in a one-dimensional, thermally perfect setting. The computational
domain is $\text{\ensuremath{\Omega}}=\left(-0.5,0.5\right)\mathrm{m}\times\left(-0.5,0.5\right)\mathrm{m}$.
The left and right boundaries are periodic, while symmetry conditions
are imposed along the top and bottom boundaries. Gmsh~\cite{Geu09}
is used to generate an unstructured triangular grid with a characteristic
cell size of $h=0.01~\mathrm{m}$. Initial $p=3$ density and pressure
fields, superimposed by the grid, are displayed in Figure~\ref{fig:thermal_bubble_2d_initial}.

\begin{figure}[h]
\subfloat[\label{fig:thermal_bubble_2d_p2_density}Density.]{\includegraphics[width=0.48\columnwidth]{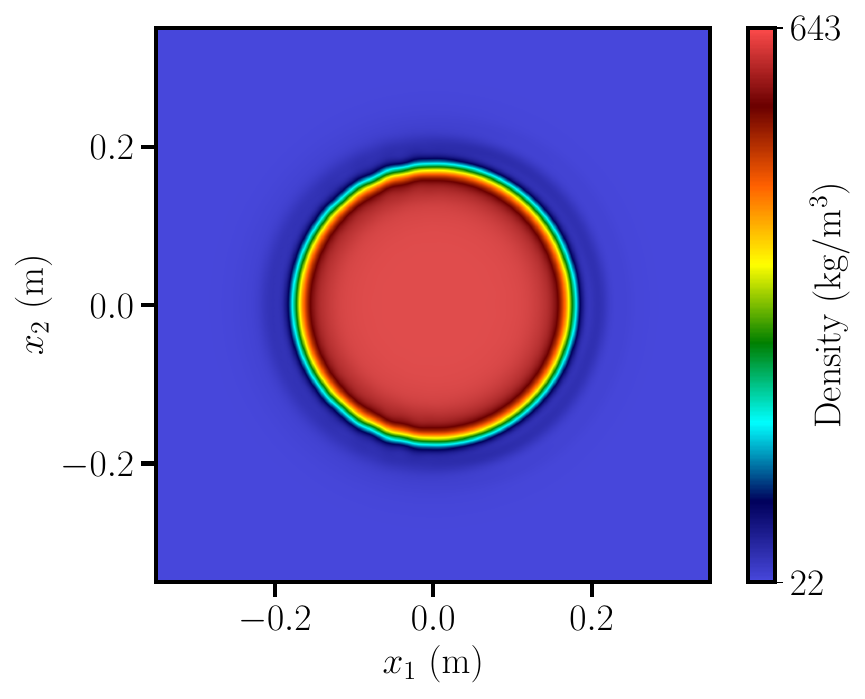}}\hfill{}\subfloat[\label{fig:thermal_bubble_2d_p2_temperature}Temperature.]{\includegraphics[width=0.48\columnwidth]{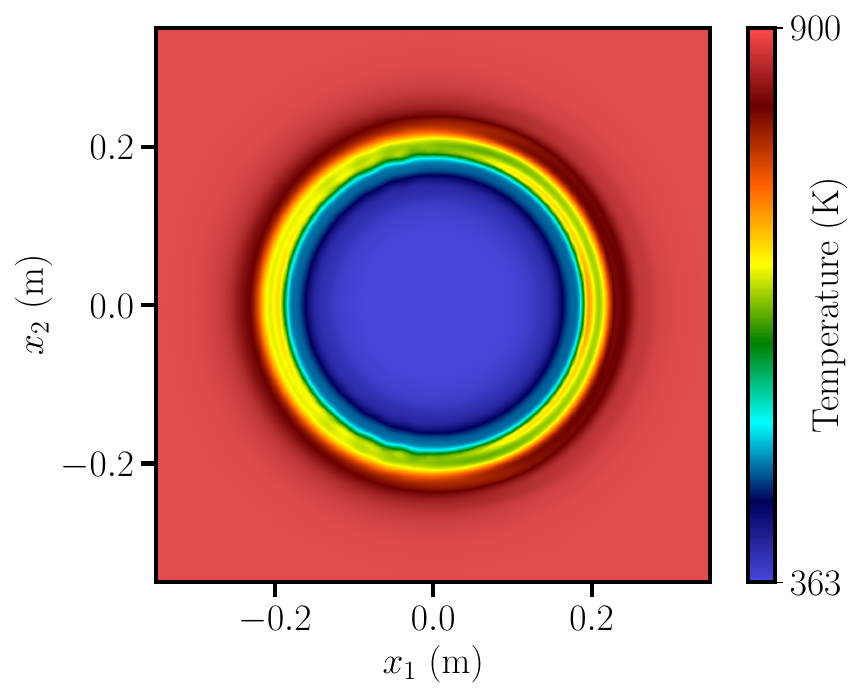}}\hfill{}\subfloat[\label{fig:thermal_bubble_2d_p2_pressure}Pressure.]{\includegraphics[width=0.48\columnwidth]{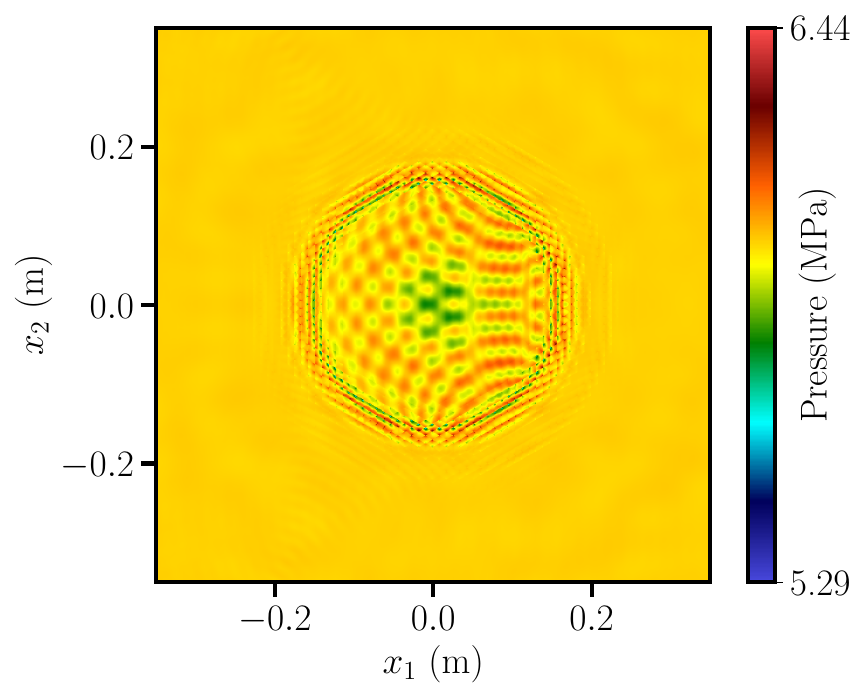}}\hfill{}\subfloat[\label{fig:thermal_bubble_2d_p2_velocity}Streamwise velocity.]{\includegraphics[width=0.48\columnwidth]{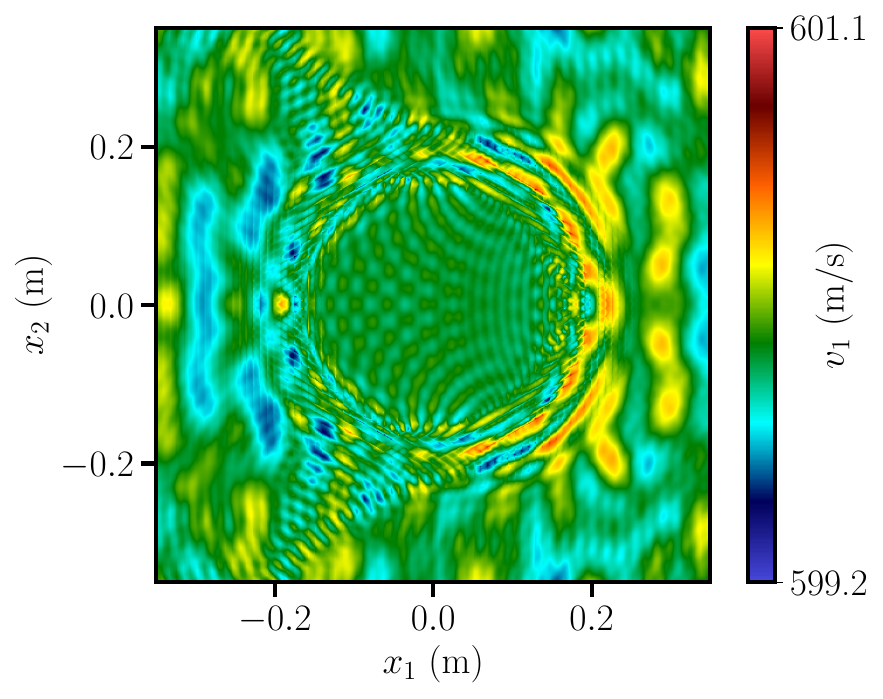}}

\caption{\label{fig:thermal_bubble_2d_p2} $p=2$ solution to two-dimensional
advection of a nitrogen/n-dodecane thermal bubble after ten periods.
The colorbar minima and maxima for the pressure and velocity fields
are the respective global minima and maxima.}
\end{figure}

\begin{figure}[h]
\subfloat[\label{fig:thermal_bubble_2d_p3_density}Density.]{\includegraphics[width=0.48\columnwidth]{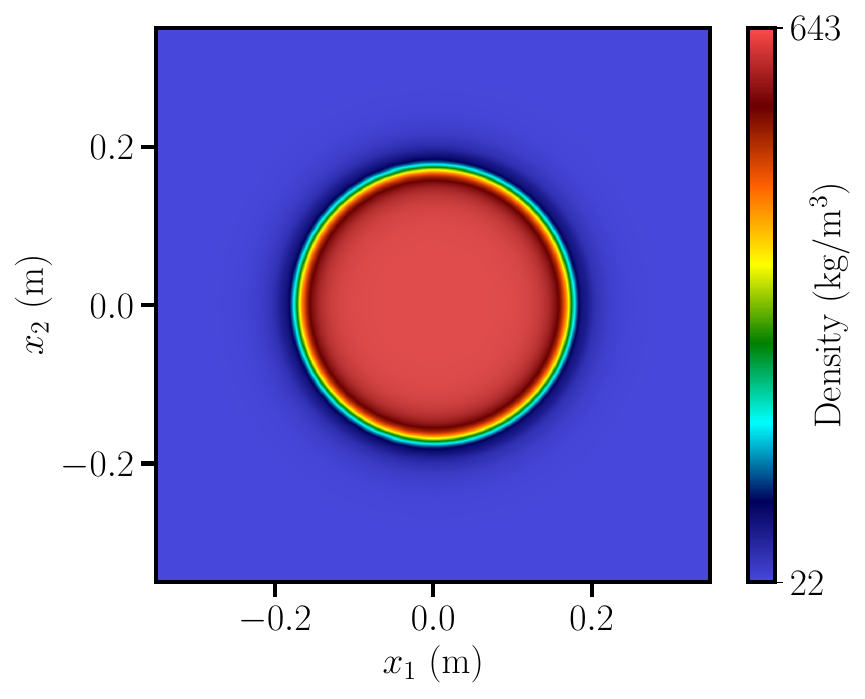}}\hfill{}\subfloat[\label{fig:thermal_bubble_2d_p3_temperature}Temperature.]{\includegraphics[width=0.48\columnwidth]{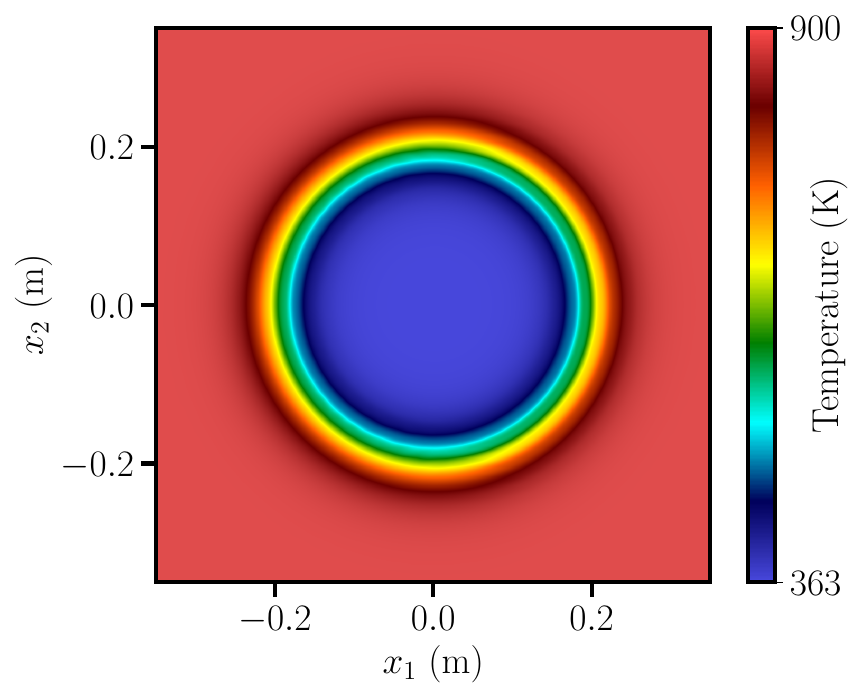}}\hfill{}\subfloat[\label{fig:thermal_bubble_2d_p3_pressure}Pressure.]{\includegraphics[width=0.48\columnwidth]{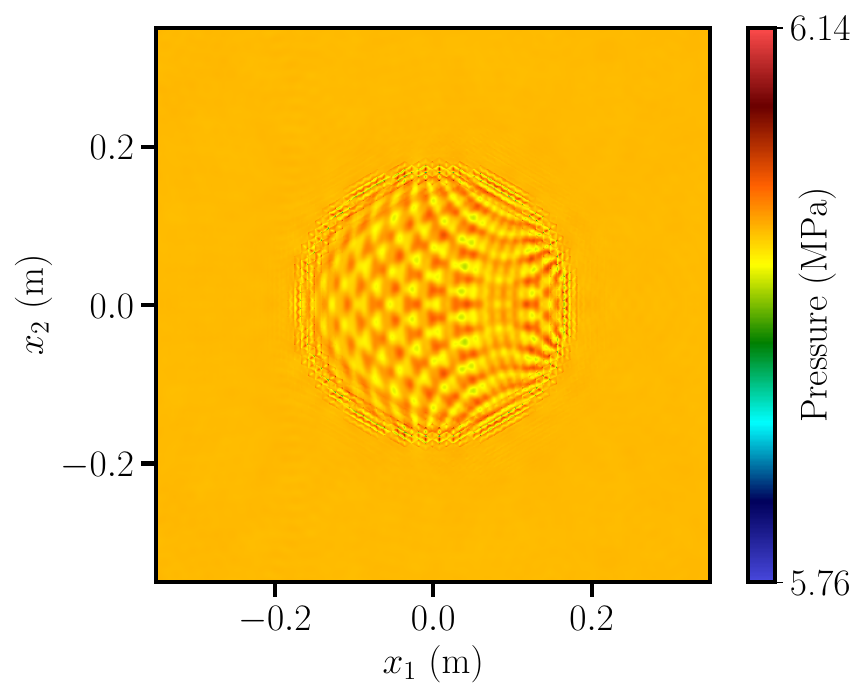}}\hfill{}\subfloat[\label{fig:thermal_bubble_2d_p3_velocity}Streamwise velocity.]{\includegraphics[width=0.48\columnwidth]{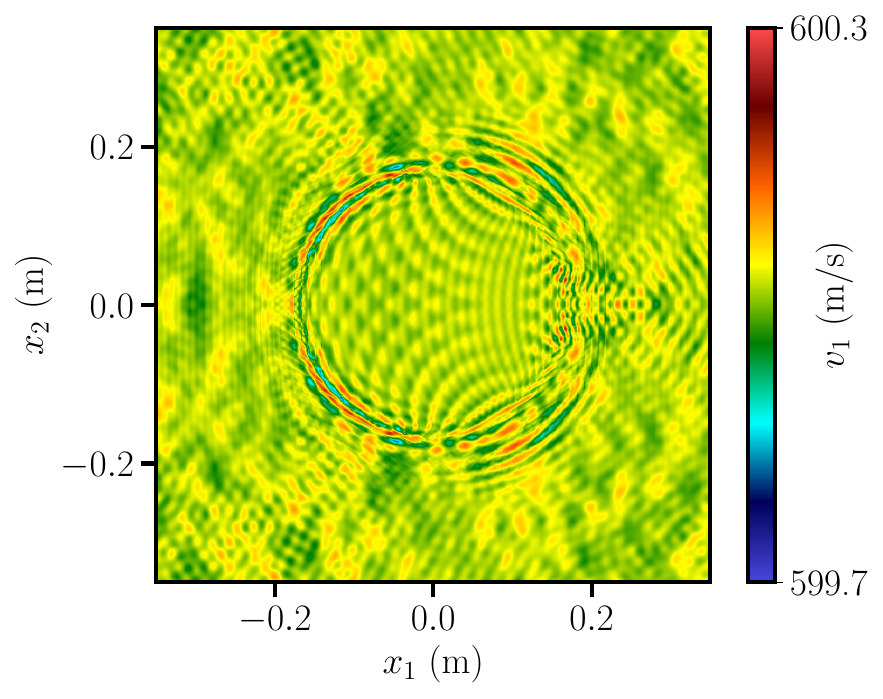}}

\caption{\label{fig:thermal_bubble_2d_p3} $p=3$ solution to two-dimensional
advection of a nitrogen/n-dodecane thermal bubble after ten periods.
The colorbar minima and maxima for the pressure and velocity fields
are the respective global minima and maxima.}
\end{figure}

As in Section~\ref{subsec:thermal_bubble}, the solutions are computed
using the overintegration~(\ref{eq:flux-projection-overintegration})
and integrated in time for ten periods using a CFL number of 0.8.
No artificial viscosity is employed. Figures~\ref{fig:thermal_bubble_2d_p2},
and~\ref{fig:thermal_bubble_2d_p3} present the final density, temperature,
pressure, and streamwise-velocity fields for $p=2$ and $p=3$ solutions,
respectively. The solver diverges for $p=1$ at these conditions.
Note that $p=1$, $p=2$, and $p=3$ with colocated integration also
fail to maintain stability.  The pressure and velocity ranges correspond
to the actual respective minima and maxima in the corresponding solutions.
Pressure oscillations are larger than in the one-dimensional case.
Only small deviations from velocity equilibrium are present. In the
$p=2$ solution, discrepancies from the initial condition near the
interfaces are observed. The $p=3$ solution more accurately preserves
the temperature and density profiles of the initial bubble and reduce
deviations from pressure equilibrium.


\begin{figure}[h]
\subfloat[\label{fig:thermal_bubble_2d_p2_density_curved}Density.]{\includegraphics[width=0.48\columnwidth]{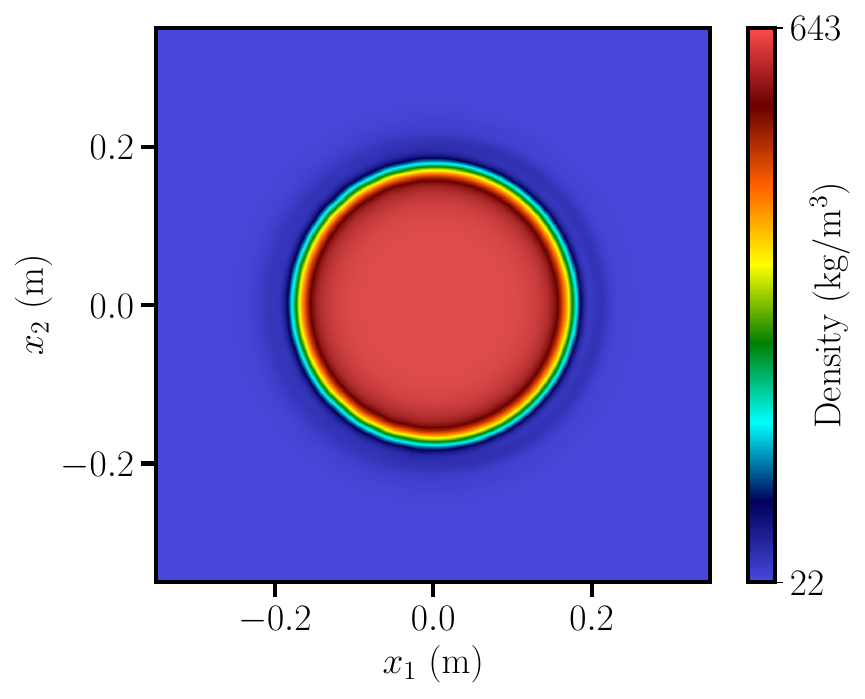}}\hfill{}\subfloat[\label{fig:thermal_bubble_2d_p2_temperature_curved}Temperature.]{\includegraphics[width=0.48\columnwidth]{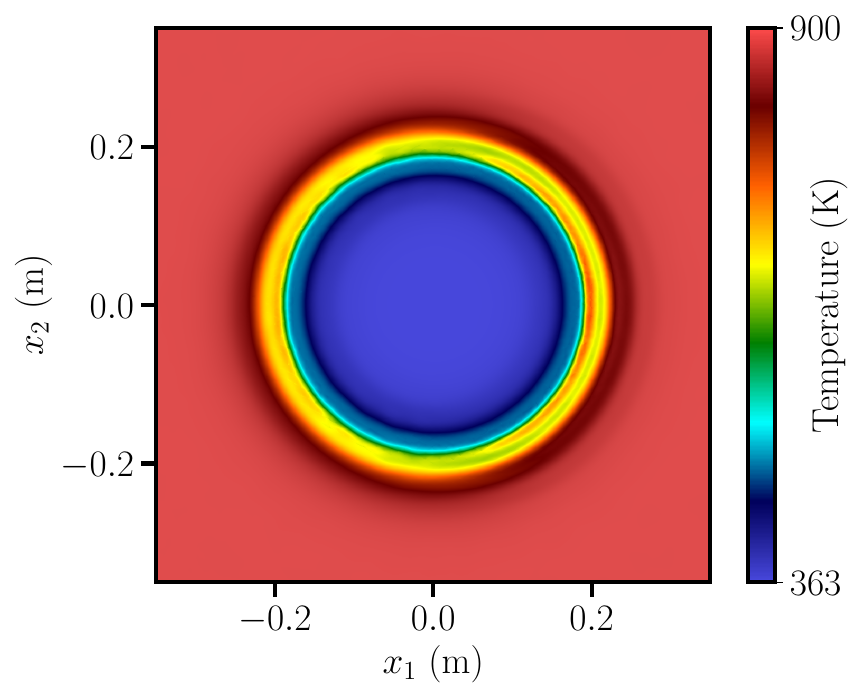}}\hfill{}\subfloat[\label{fig:thermal_bubble_2d_p2_pressure_curved}Pressure.]{\includegraphics[width=0.48\columnwidth]{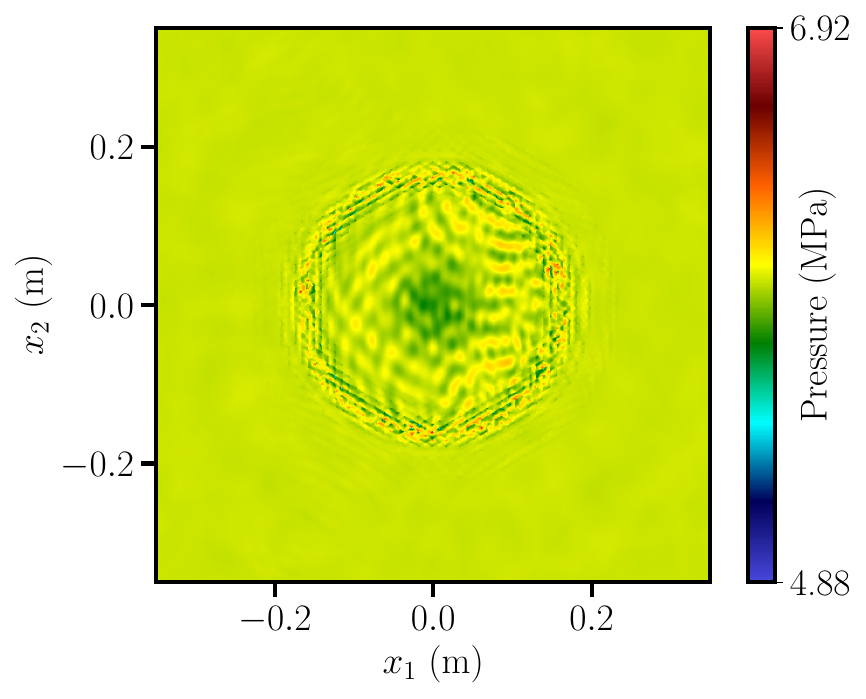}}\hfill{}\subfloat[\label{fig:thermal_bubble_2d_p2_velocity_curved}Streamwise velocity.]{\includegraphics[width=0.48\columnwidth]{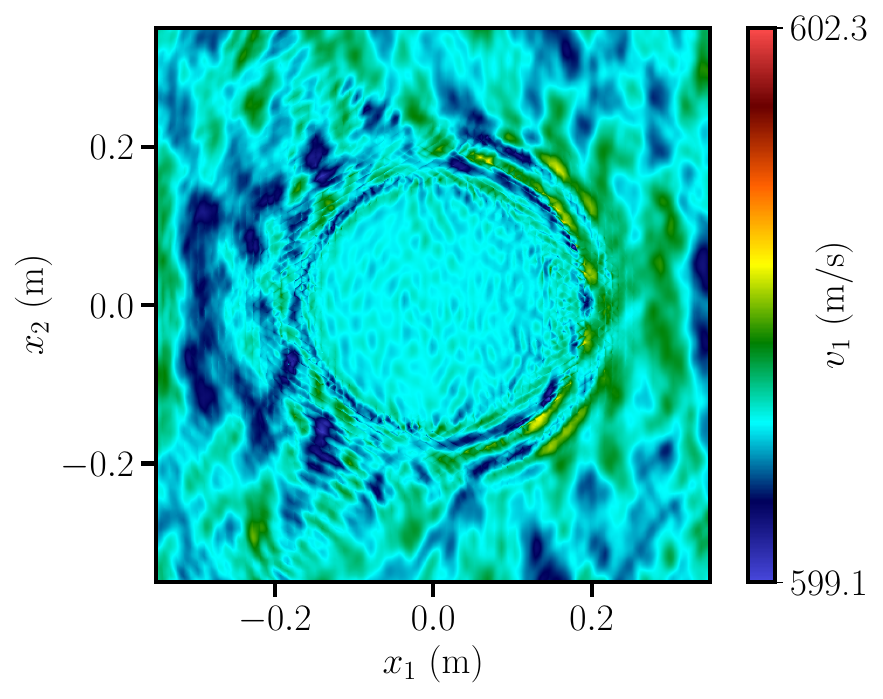}}

\caption{\label{fig:thermal_bubble_2d_p2_curved} $p=2$ solution to two-dimensional
advection of a nitrogen/n-dodecane thermal bubble after ten periods
on a curved grid. The colorbar minima and maxima for the pressure
and velocity fields are the respective global minima and maxima.}
\end{figure}

\begin{figure}[h]
\subfloat[\label{fig:thermal_bubble_2d_p3_density_curved}Density.]{\includegraphics[width=0.48\columnwidth]{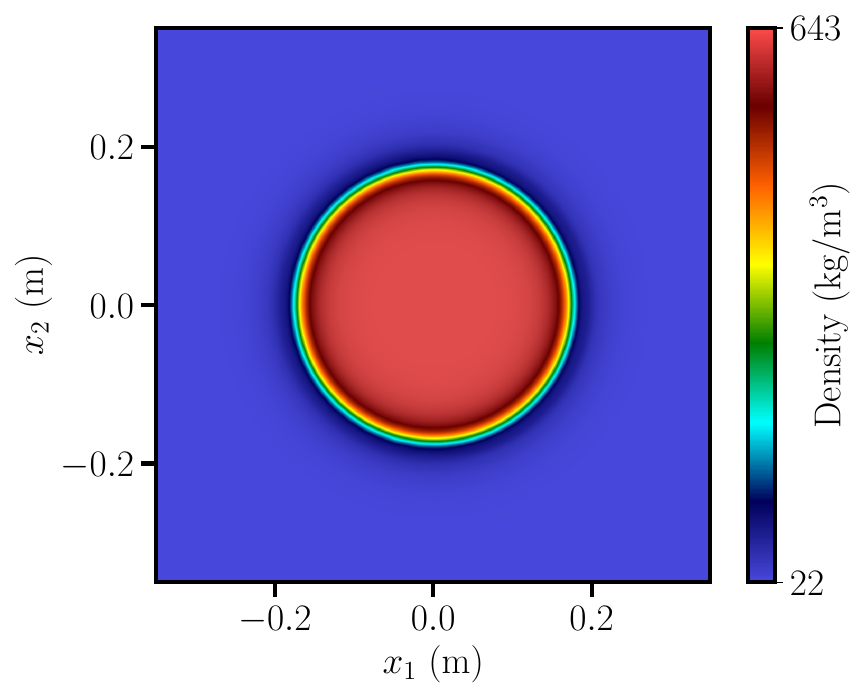}}\hfill{}\subfloat[\label{fig:thermal_bubble_2d_p3_temperature_curved}Temperature.]{\includegraphics[width=0.48\columnwidth]{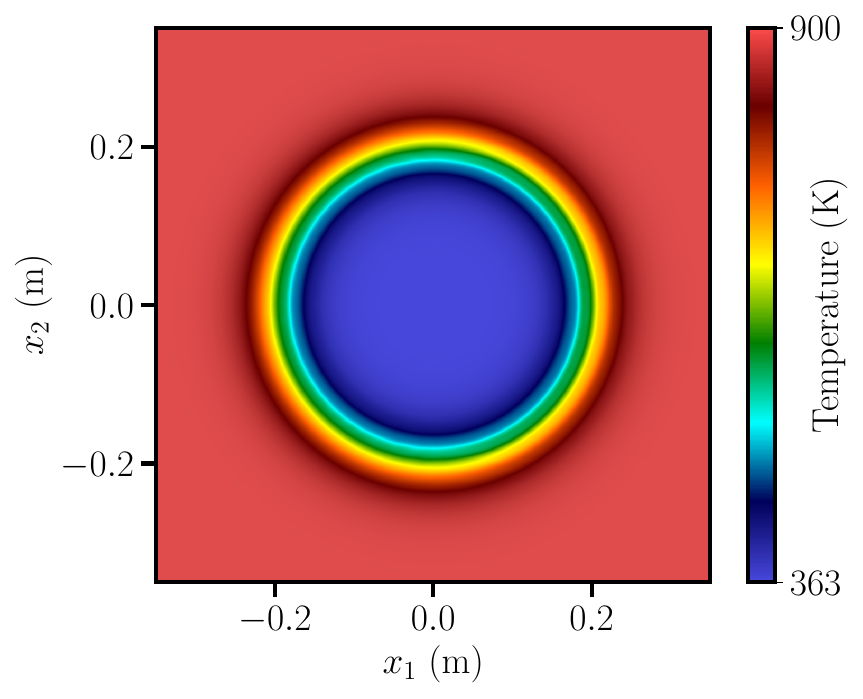}}\hfill{}\subfloat[\label{fig:thermal_bubble_2d_p3_pressure_curved}Pressure.]{\includegraphics[width=0.48\columnwidth]{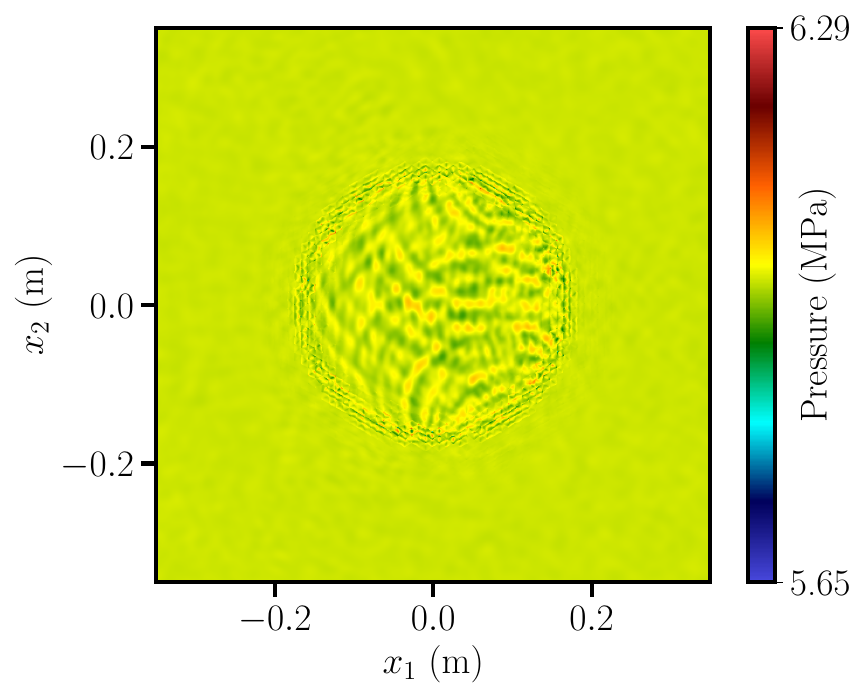}}\hfill{}\subfloat[\label{fig:thermal_bubble_2d_p3_velocity_curved}Streamwise velocity.]{\includegraphics[width=0.48\columnwidth]{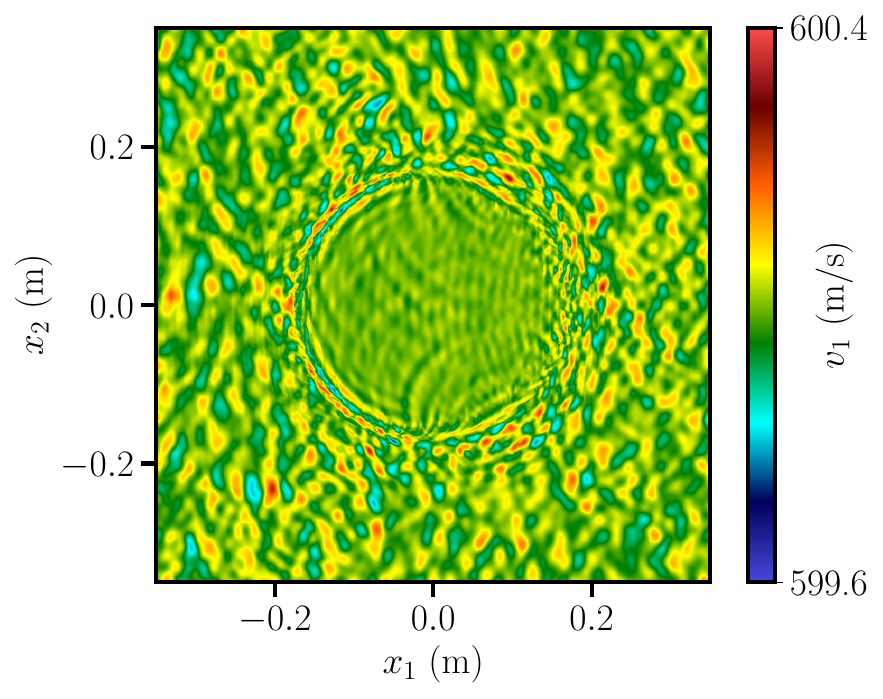}}

\caption{\label{fig:thermal_bubble_2d_p3_curved} $p=3$ solution to two-dimensional
advection of a nitrogen/n-dodecane thermal bubble after ten periods
on a curved grid. The colorbar minima and maxima for the pressure
and velocity fields are the respective global minima and maxima.}
\end{figure}

Finally, we recompute the $p=2$ and $p=3$ cases using curved elements
of quadratic geometric order. To generate the curved grid, high-order
geometric nodes are inserted at the midpoints of the vertices of each
element. At interior edges, the midpoint nodes are randomly perturbed
by a distance up to $0.03h$. Figures~\ref{fig:thermal_bubble_2d_p2_curved}
and~\ref{fig:thermal_bubble_2d_p3_curved} display the $p=2$ and
$p=3$ results, respectively. The deviations from pressure and velocity
equilibrium are slightly larger than for the straight-sided grid,
but the density and temperature fields remain accurately predicted,
suggesting that the employed overintegration strategy can maintain
solution stability on curved grids.

\subsection{Two-dimensional n-dodecane jet\label{subsec:Two-dimensional-n-dodecane-jet}}

Next, we simulate two-dimensional injection of an n-dodecane jet into
a quiescent nitrogen environment. Since phase separation is outside
the scope of this work, we target the supercritical conditions considered
by Rodriguez et al.~\cite{Rod19}. The computational domain is a
$5\text{ mm}\times2.5\text{ mm}$ rectangular chamber, which is discretized
by a fully unstructured triangular grid with a characteristic cell
size of $h=20\text{ \ensuremath{\mu}m}$. The total number of cells
is 72,098. The domain is initialized with nitrogen at a pressure of
11.1 MPa and density of $37\text{ kg/m}^{3}$. The top and bottom
boundaries are periodic, and the chamber pressure is specified at
the outlet. The n-dodecane jet is injected at a streamwise velocity,
pressure, and density of 200 m/s, 11.1 MPa, and $400\text{ kg/m}^{3}$,
respectively, through a 0.1 mm-wide nozzle inlet located at the center
of the left boundary. The remainder of the left boundary is a slip
wall. No smoothing of the inflow is performed.  Note that Rodriguez
et al.~\cite{Rod19} simulated this case with the PC-SAFT equation
of state and viscous effects.

\begin{figure}[h]
\subfloat[\label{fig:dodecane_jet_2d_p1_density}Density.]{\includegraphics[width=0.48\columnwidth]{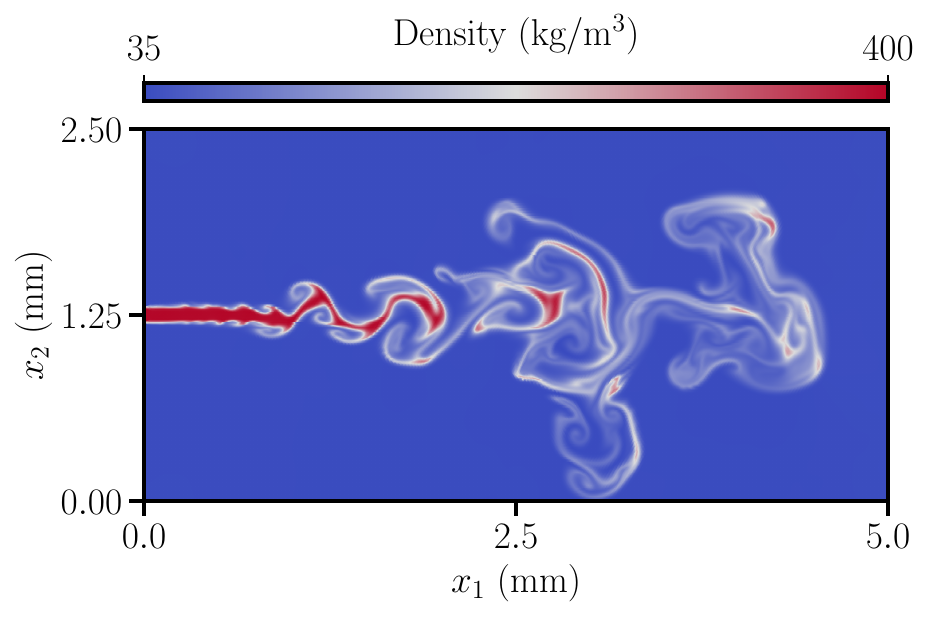}}\hfill{}\subfloat[\label{fig:dodecane_jet_2d_p1_temperature}Temperature.]{\includegraphics[width=0.48\columnwidth]{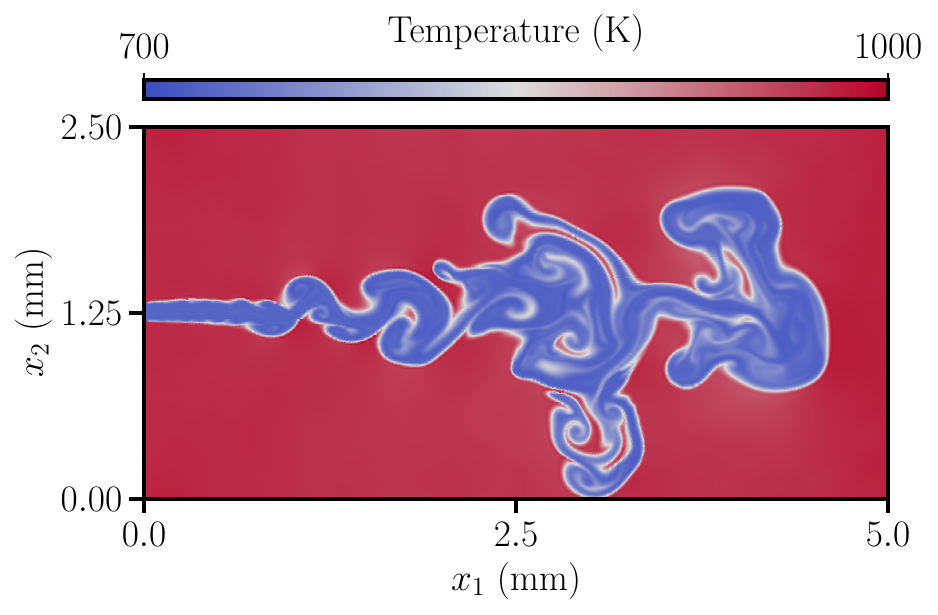}}\hfill{}\subfloat[\label{fig:dodecane_jet_2d_p1_pressure}Pressure.]{\includegraphics[width=0.48\columnwidth]{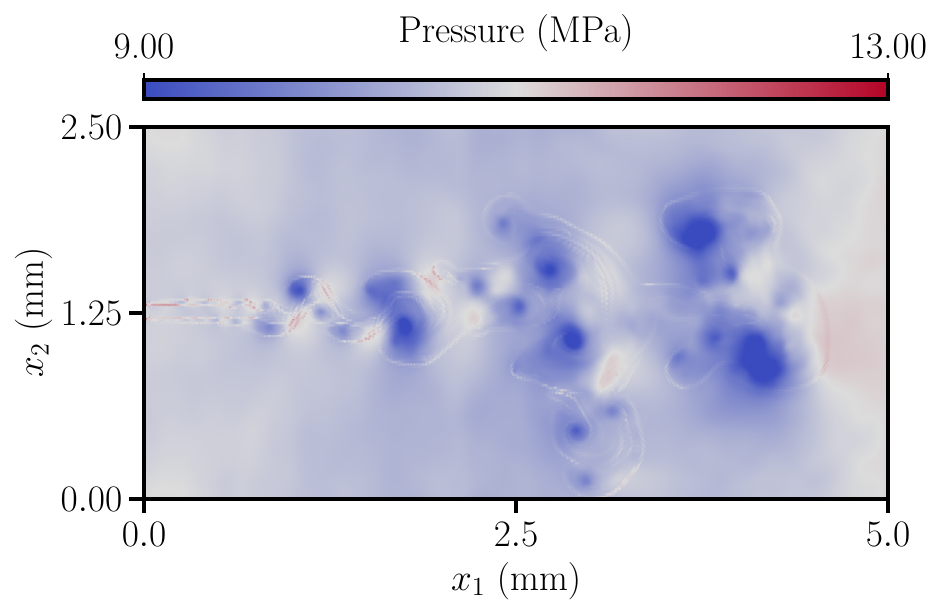}}\hfill{}\subfloat[\label{fig:dodecane_jet_2d_p1_velocity}Streamwise velocity.]{\includegraphics[width=0.48\columnwidth]{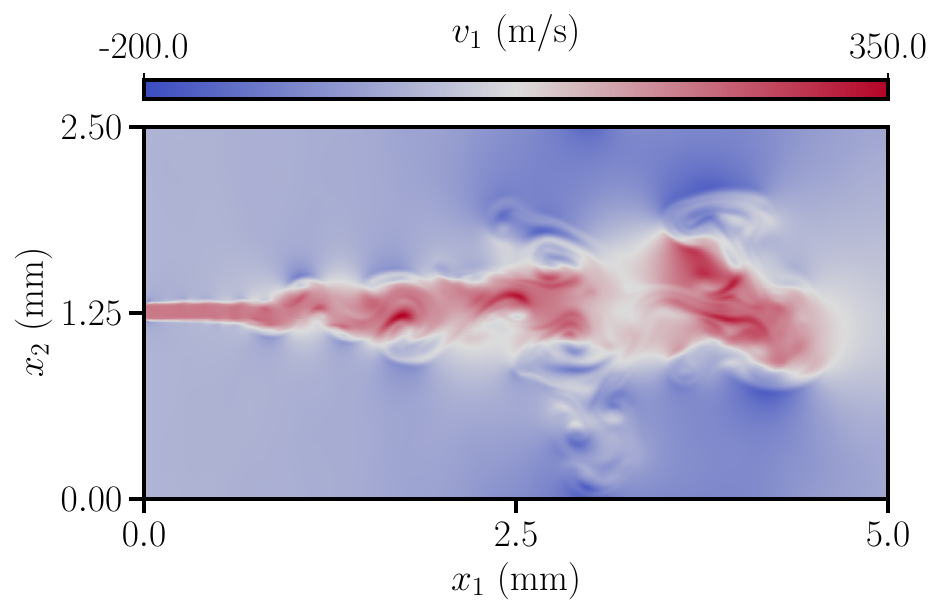}}

\caption{\label{fig:dodecane_jet_2d_p1} $p=1$ solution to two-dimensional
injection of an n-dodecane jet into a quiescent nitrogen environment
at $t=\text{35 \ensuremath{\mu}s}$.}
\end{figure}

\begin{figure}[h]
\subfloat[\label{fig:dodecane_jet_2d_p2_density}Density.]{\includegraphics[width=0.48\columnwidth]{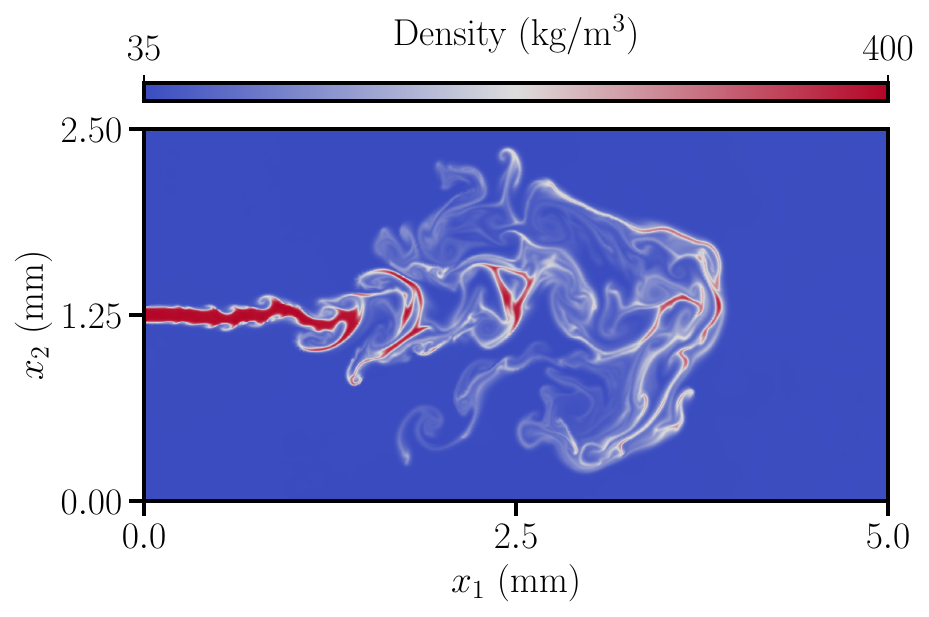}}\hfill{}\subfloat[\label{fig:dodecane_jet_2d_p2_temperature}Temperature.]{\includegraphics[width=0.48\columnwidth]{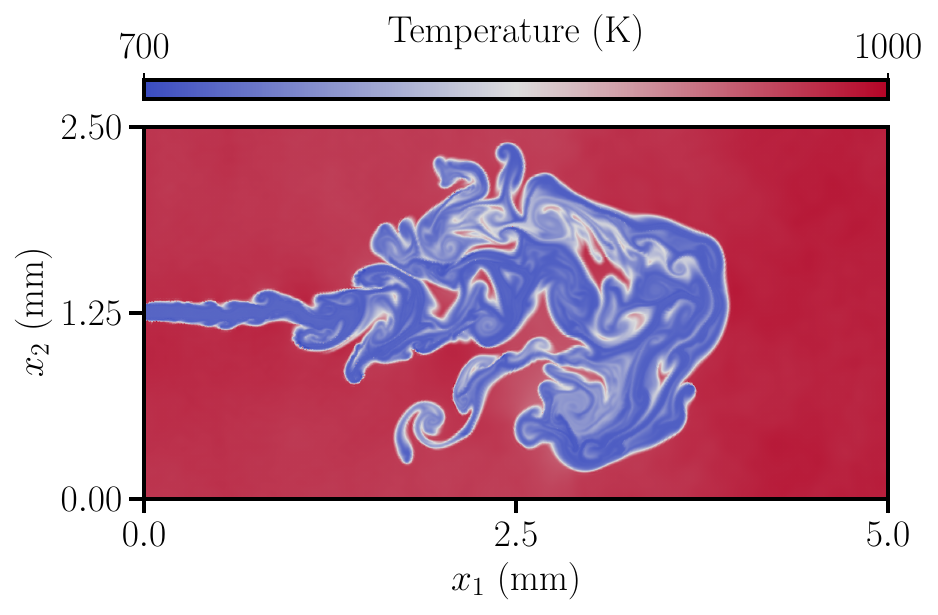}}\hfill{}\subfloat[\label{fig:dodecane_jet_2d_p2_pressure}Pressure.]{\includegraphics[width=0.48\columnwidth]{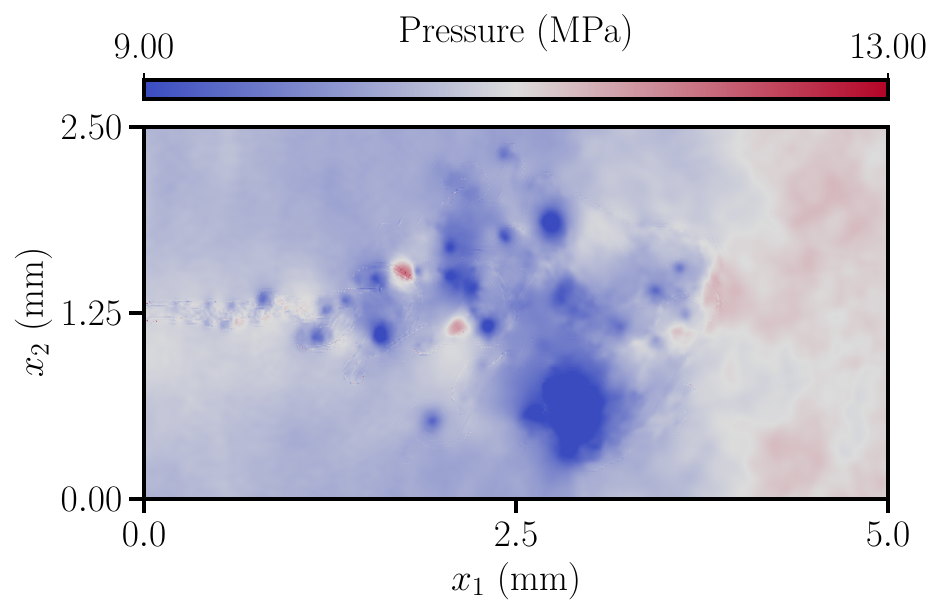}}\hfill{}\subfloat[\label{fig:dodecane_jet_2d_p2_velocity}Streamwise velocity.]{\includegraphics[width=0.48\columnwidth]{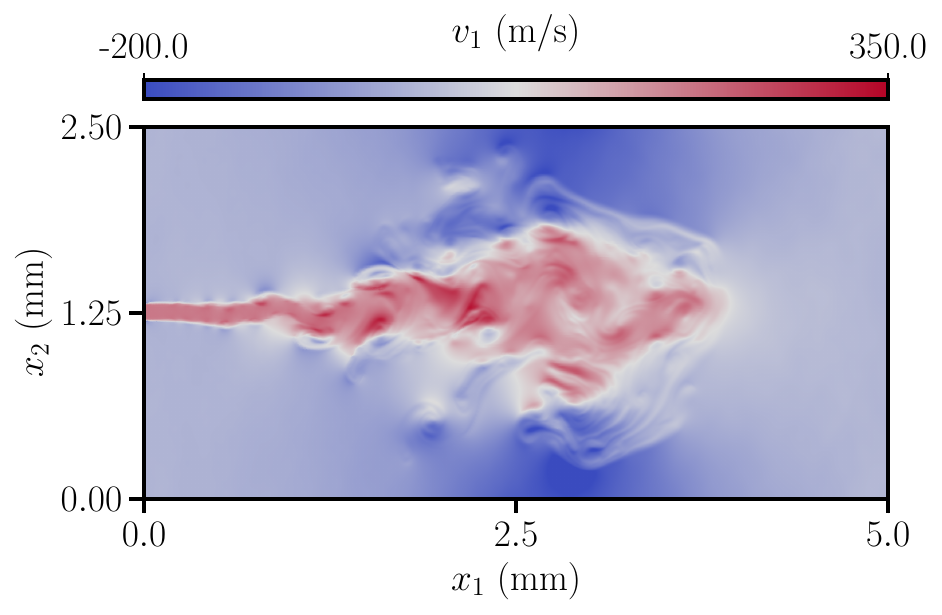}}

\caption{\label{fig:dodecane_jet_2d_p2} $p=2$ solution to two-dimensional
injection of an n-dodecane jet into a quiescent nitrogen environment
at $t=\text{35 \ensuremath{\mu}s}$.}
\end{figure}

\begin{figure}[h]
\subfloat[\label{fig:dodecane_jet_2d_p3_density}Density.]{\includegraphics[width=0.48\columnwidth]{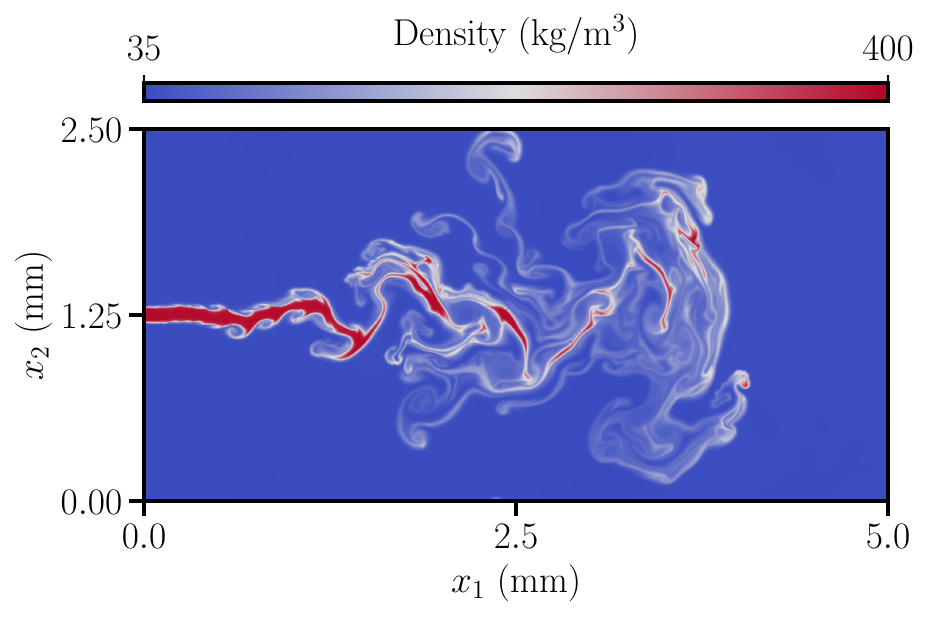}}\hfill{}\subfloat[\label{fig:dodecane_jet_2d_p3_temperature}Temperature.]{\includegraphics[width=0.48\columnwidth]{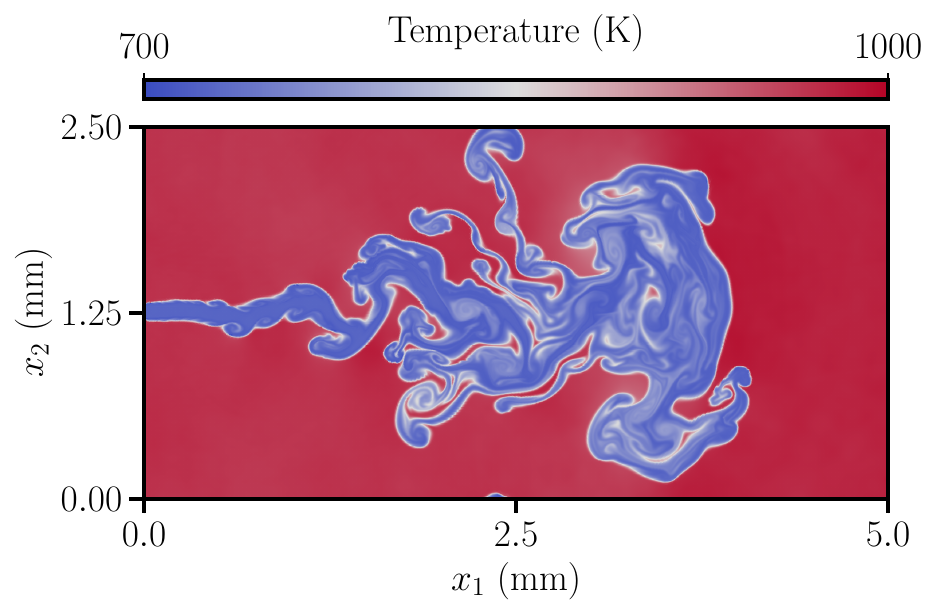}}\hfill{}\subfloat[\label{fig:dodecane_jet_2d_p3_pressure}Pressure.]{\includegraphics[width=0.48\columnwidth]{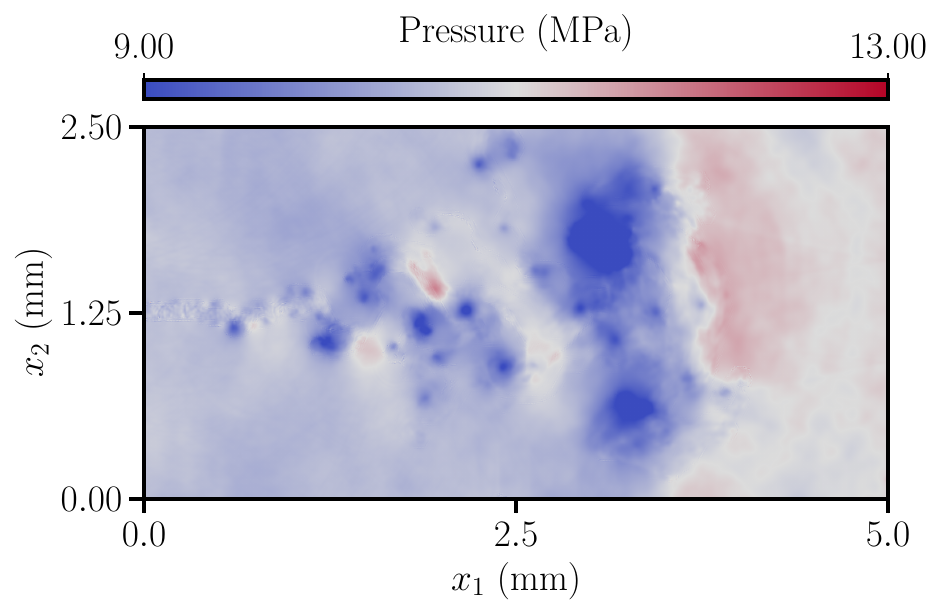}}\hfill{}\subfloat[\label{fig:dodecane_jet_2d_p3_velocity}Streamwise velocity.]{\includegraphics[width=0.48\columnwidth]{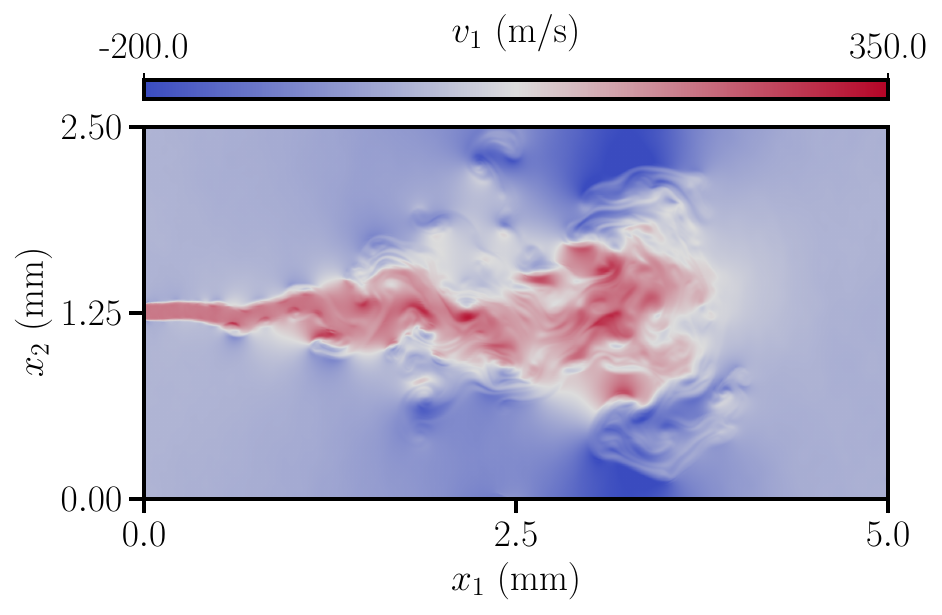}}

\caption{\label{fig:dodecane_jet_2d_p3} $p=3$ solution to two-dimensional
injection of an n-dodecane jet into a quiescent nitrogen environment
at $t=\text{35 \ensuremath{\mu}s}$.}
\end{figure}

We compute $p=1$, $p=2$, and $p=3$ solutions up to $t=\text{35 \ensuremath{\mu}s}$
with a CFL number of 0.6. Figures~\ref{fig:dodecane_jet_2d_p1},~\ref{fig:dodecane_jet_2d_p2},
and~\ref{fig:dodecane_jet_2d_p3} display the density, temperature,
pressure, and streamwise-velocity fields for $p=1$, $p=2$, and $p=3$,
respectively. Our results are qualitatively similar to those by Rodriguez
et al.~\cite{Rod19}. At these supercritical conditions, gas-like
mixing is observed, instead of classical liquid-jet breakup into ligaments
and droplets~\cite{Che03}. The boundary between the dense jet core
and the ambient nitrogen appears as a sharp yet smooth transition.
The dense core is shorter and thicker in the $p=2$ and $p=3$ solutions
than in the $p=1$ solution. Furthermore, the $p=2$ and $p=3$ solutions
exhibit more vortical features and ``finger-like'' structures in
the downstream region of the jet. Although spurious pressure oscillations
can be observed in all cases, they remain relatively small and do
not cause solver divergence. The density, temperature, and velocity
fields are generally free from spurious artifacts. 


\subsection{Three-dimensional n-dodecane jet\label{subsec:Three-dimensional-n-dodecane-jet}}

The final test case is a three-dimensional extension of that in Section~\ref{subsec:Two-dimensional-n-dodecane-jet},
namely injection of a three-dimensional n-dodecane jet into a nitrogen
environment. The computational domain is a cylindrical chamber of
length 5 mm and radius 1.25 mm. The axis of the cylindrical domain
is the $x$-axis. As in Section~\ref{subsec:Two-dimensional-n-dodecane-jet},
the domain is initialized with nitrogen at a pressure of 11.1 MPa
and density of $37\text{ kg/m}^{3}$. Slip-wall conditions are imposed
along the walls of the cylinder, and the chamber pressure is specified
at the outflow. The n-dodecane jet is injected at a streamwise velocity,
pressure, and density of 200 m/s, 11.1 MPa, and $400\text{ kg/m}^{3}$,
respectively, through a cylindrical nozzle inlet of radius 0.05 mm.
The axis of the nozzle (not included in the computational domain)
is the $x$-axis. 

Gmsh~\cite{Geu09} is used to generate an unstructured mesh with
approximately 19 million tetrahedral cells. The characteristic cell
size is specified to be $h=10\text{ \ensuremath{\mu}m}$ within the
truncated cone
\[
\mathsf{C}=\left\{ \sqrt{x_{2}^{2}+x_{3}^{2}}\leq R_{1}+\frac{R_{2}-R_{1}}{L}x_{1}|0\leq x_{1}\leq L\right\} ,
\]
where $R_{1}=0.075\text{ mm}$, $R_{2}=0.95\text{ mm}$, and $L=4\text{ mm}$.
The mesh transitions to a characteristic cell size of $h=0.5\text{ mm}$
at the domain boundaries. 

\begin{figure}[h]
\subfloat[\label{fig:dodecane_jet_3d_p1_density}Density.]{\includegraphics[width=0.48\columnwidth]{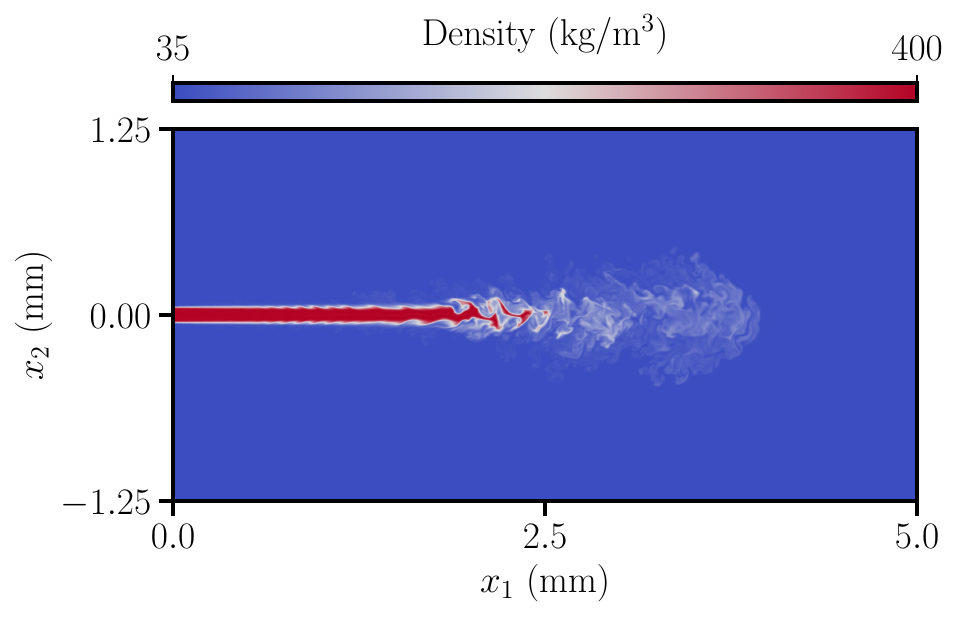}}\hfill{}\subfloat[\label{fig:dodecane_jet_3d_p1_temperature}Temperature.]{\includegraphics[width=0.48\columnwidth]{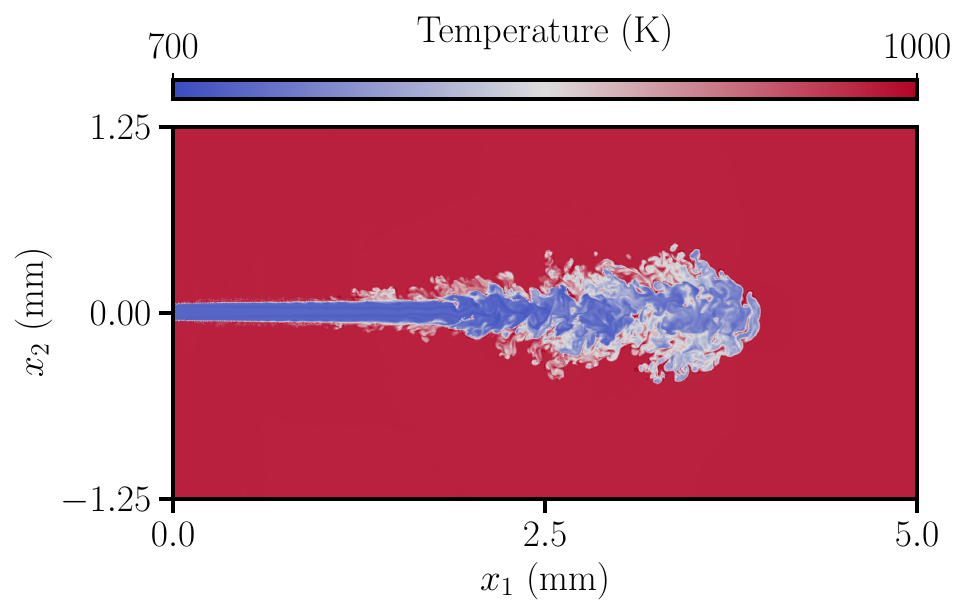}}\hfill{}\subfloat[\label{fig:dodecane_jet_3d_p1_pressure}Pressure.]{\includegraphics[width=0.48\columnwidth]{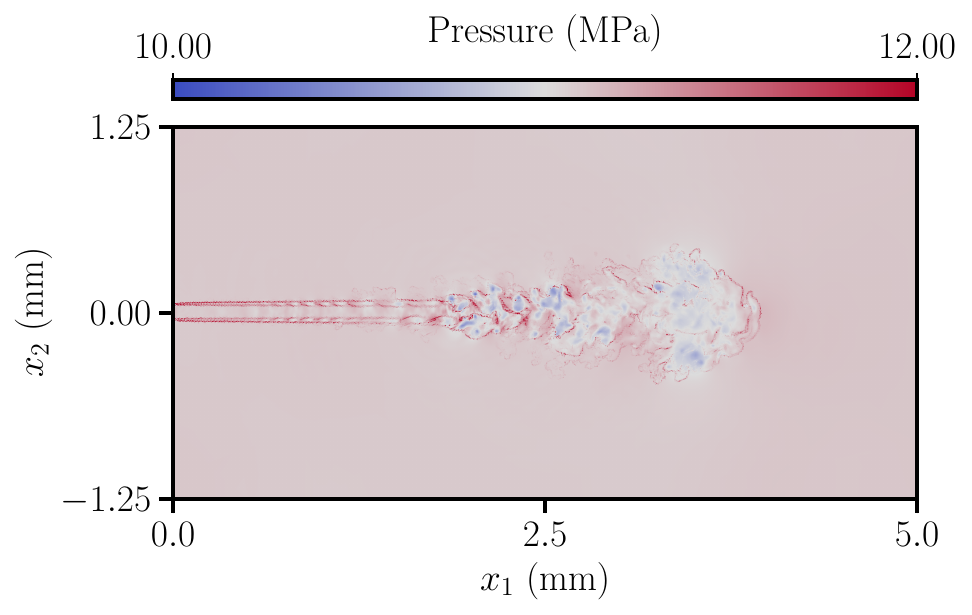}}\hfill{}\subfloat[\label{fig:dodecane_jet_3d_p1_velocity}Streamwise velocity.]{\includegraphics[width=0.48\columnwidth]{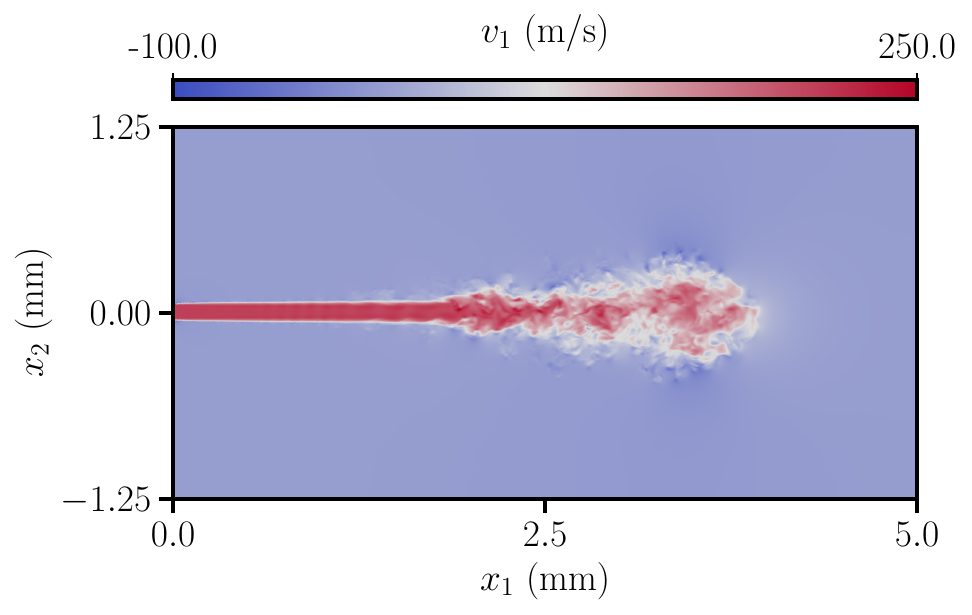}}

\caption{\label{fig:dodecane_jet_3d_p1} $p=1$ solution to three-dimensional
injection of an n-dodecane jet into a quiescent nitrogen environment
at $t=\text{35 \ensuremath{\mu}s}$.}
\end{figure}

\begin{figure}[h]
\subfloat[\label{fig:dodecane_jet_3d_p2_density}Density.]{\includegraphics[width=0.48\columnwidth]{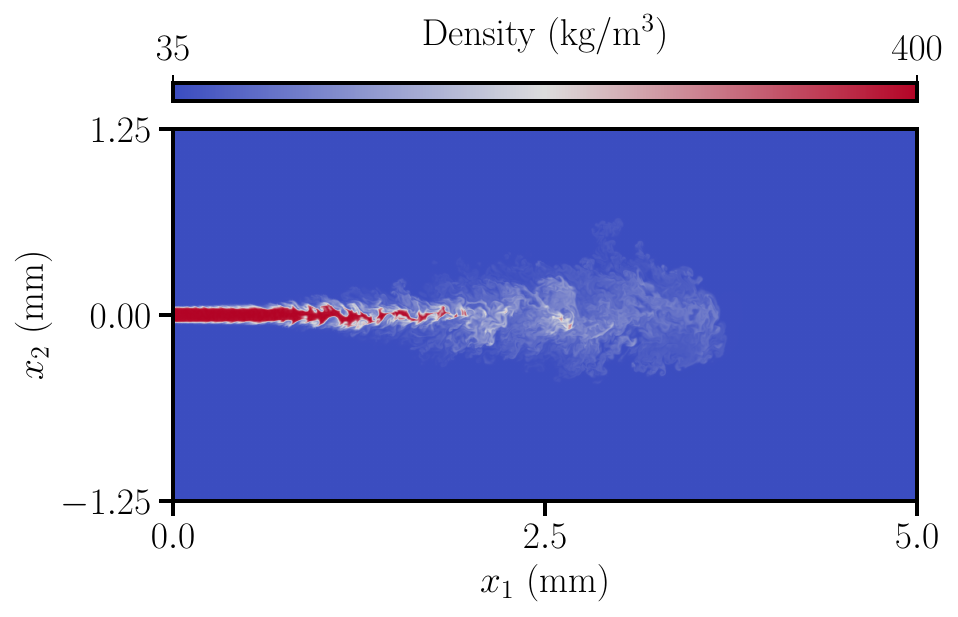}}\hfill{}\subfloat[\label{fig:dodecane_jet_3d_p2_temperature}Temperature.]{\includegraphics[width=0.48\columnwidth]{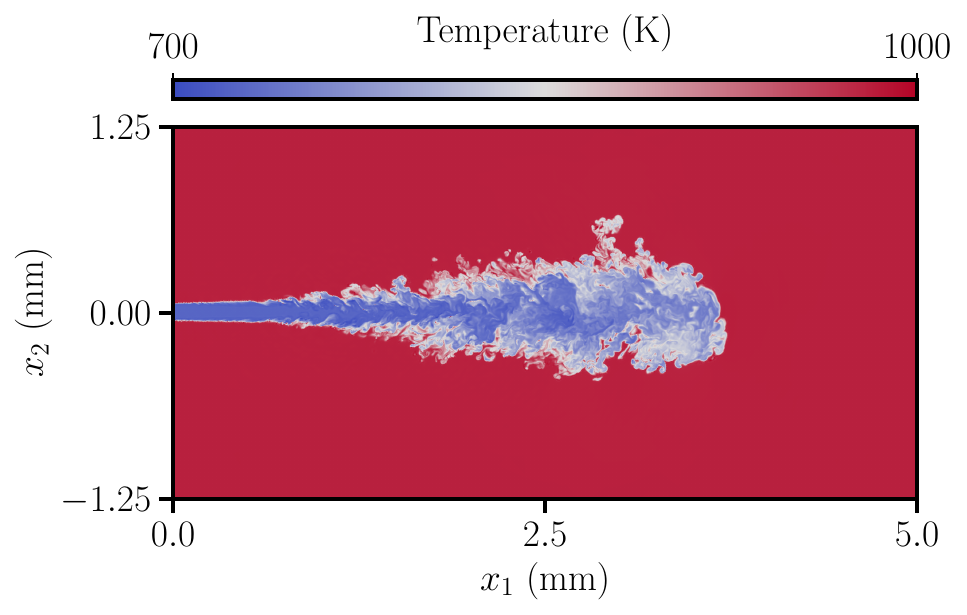}}\hfill{}\subfloat[\label{fig:dodecane_jet_3d_p2_pressure}Pressure.]{\includegraphics[width=0.48\columnwidth]{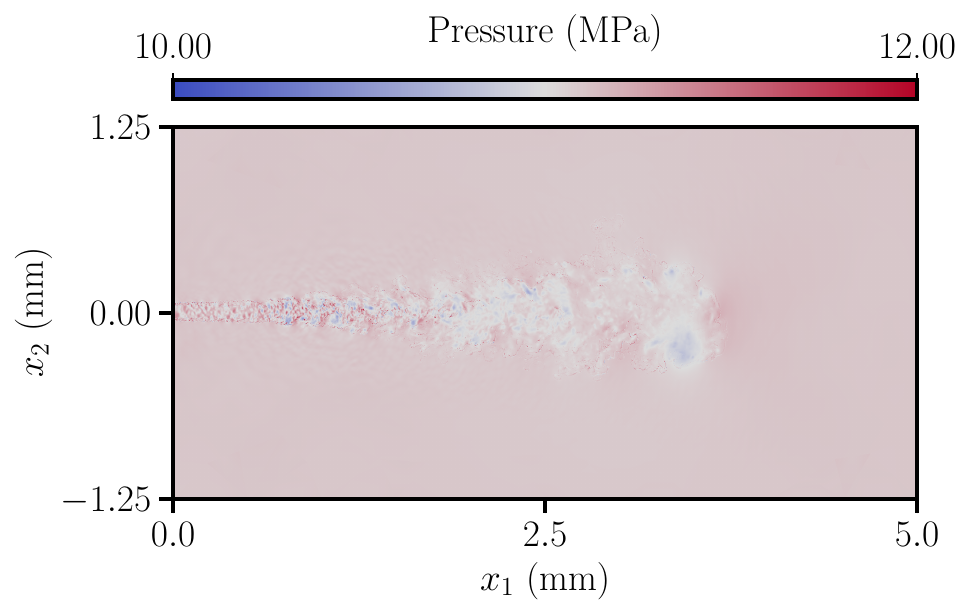}}\hfill{}\subfloat[\label{fig:dodecane_jet_3d_p2_velocity}Streamwise velocity.]{\includegraphics[width=0.48\columnwidth]{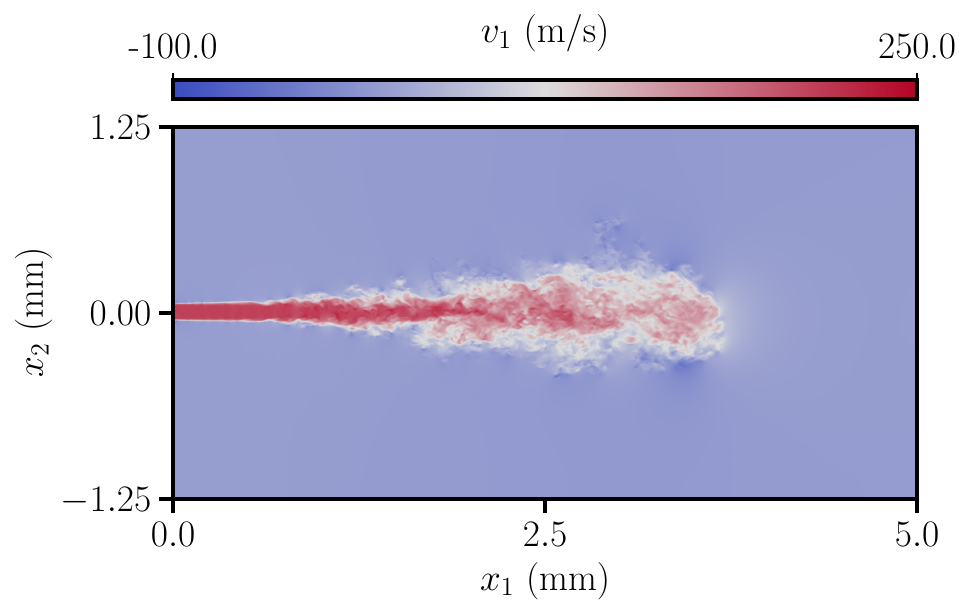}}

\caption{\label{fig:dodecane_jet_3d_p2} $p=2$ solution to three-dimensional
injection of an n-dodecane jet into a quiescent nitrogen environment
at $t=\text{35 \ensuremath{\mu}s}$.}
\end{figure}

Figure~\ref{fig:dodecane_jet_3d_p1} presents the density, temperature,
pressure, and streamwise-velocity fields for a $p=1$ solution at
$t=\text{35 \ensuremath{\mu}s}$. In order to reduce nonphysical instabilities
at startup, the injection velocity is linearly ramped up from 0 m/s
at $t=0$ to 200 m/s at $t=\text{3 \ensuremath{\mu}s}$. Compared
to the two-dimensional results in Section~\ref{subsec:Two-dimensional-n-dodecane-jet},
the three-dimensional dense jet core here is significantly narrower,
and the finger-like structures are finer. Note, however, that periodicity
is imposed at the top and bottom walls in Section~\ref{subsec:Two-dimensional-n-dodecane-jet}.
The boundary between the jet and the ambient nitrogen is sharp yet
smooth. Figure~\ref{fig:dodecane_jet_3d_p2} shows the density, temperature,
pressure, and streamwise-velocity fields for a $p=1$ solution at
$t=\text{35 \ensuremath{\mu}s}$. The $p=2$ solution is started from
the $p=1$ solution at approximately $t=\text{3.35 \ensuremath{\mu}s}$.
Some spurious oscillations can be observed in both cases, although
those in the $p=1$ solution are more prominent at the interface.
In addition, compared to the $p=1$ solution, breakup of the jet occurs
further upstream in the $p=2$ solution, and the small-scale flow
features are better resolved. Nevertheless, the $p=1$ and $p=2$
solutions are overall qualitatively similar. These results demonstrate
the ability of the employed DG formulation to compute complex supercritical
flows in three dimensions. 

\section{Concluding remarks}

In this work, we presented a conservative DG methodology for the simulation
of transcritical/supercritical, real-fluid flows without phase change.
The methodology builds on a DG formulation originally developed to
simulate multicomponent flows involving mixtures of thermally perfect
gases~\cite{Joh20_2,Chi23_short}. A key feature of the presented
methodology is an overintegration strategy that employs an $L^{2}$-projection
of a set of intermediate variables onto the finite element test space
when evaluating the flux, similar to techniques previously discussed
in~\cite{Fra16} and~\cite{Ban23}. In the context of mixtures of
thermally perfect gases, this overintegration strategy was found to
effectively maintain approximate pressure equilibrium at smooth contact
interfaces~\cite{Chi23_short} without relying on dissipative stabilization
in the form of artificial viscosity or limiting. Here, the ability
of this strategy to maintain pressure equilibrium and solution stability
in the presence of contact interfaces even with the added nonlinearities
associated with a cubic equation of state and thermodynamic departure
functions was of particular interest.

We applied the formulation to five test cases. In the first, which
involved the advection of a sinusoidal density wave, optimal convergence
was verified. In the second, we considered one-dimensional advection
of a nitrogen/n-dodecane thermal bubble. In the thermally perfect
case, both colocated integration and the employed overintegration
strategy maintained solution stability over ten advection periods.
However, with real-fluid effects accounted for, the former failed
to prevent solver failure in nearly all considered simulations. In
contrast, the latter maintained solution stability and small deviations
from pressure equilibrium in all cases except the coarsest one. Increased
resolution led to improved preservation of pressure equilibrium. The
third test case involved a two-dimensional version of the previous
configuration. Although spurious pressure oscillations were noticeably
larger than in the one-dimensional case, the methodology was able
to maintain solution stability on unstructured grids, including a
grid with curved elements. Deviations from velocity equilibrium were
generally small. The final test cases comprised two- and three-dimensional
injection of an n-dodecane jet into a quiescent nitrogen environment,
demonstrating the ability of the employed DG formulation to compute
complex supercritical flows on unstructured simplicial grids.

Although these results are encouraging, there are still issues that
should be addressed in future work. For example, the development of
provably nonlinearly stable (e.g., positivity-preserving and entropy-stable)
schemes for supercritical and transcritical flows would significantly
improve robustness. However, the complex thermodynamics associated
with this regime (especially in the case of mixtures) can introduce
challenges such as loss of satisfaction of mathematical properties
required by some conventional positivity-preserving schemes~\cite{Zha10,Chi22,Chi22_2}.
Recent work by Clayton et al.~\cite{Cla23} represents an encouraging
step in this direction. In addition, entropy-stable schemes rely on
a mathematical entropy function that is convex with respect to the
state~\cite{Har83_3,Che17,Cha18_2}, but common entropy functions
that are valid in the ideal-gas case may lose convexity under certain
conditions when real-fluid models are employed~\cite{Gio13}. Another
avenue for future work is the development of mathematically pressure-equilibrium-preserving
schemes based on the conservative variables that do not rely on frozen
thermodynamics (unlike the double-flux strategy). Recent work in this
area has led to a number of promising formulations. For example, Bernades
et al.~\cite{Ber23} devised a kinetic-energy- and pressure-equilibrium-preserving
finite-difference/finite-volume method, although total-energy conservation
is still sacrificed. Terashima et al.~\cite{Ter24} developed a method
that is fully conservative and approximately pressure-equilibrium-preserving
(note that the DG formulation here employs techniques that can reduce
spurious pressure oscillations, but the pressure-equilibrium condition
is not specifically invoked). However, mathematically guaranteeing
both total-energy conservation and pressure-equilibrium preservation
has remained elusive. Finally, in order to enable consideration of
phase change in the transcritical regime, we will incorporate vapor-liquid-equilibrium
calculations~\cite{Ma19_2,Zha24} into the physical model.

\section*{Acknowledgments}

This work is sponsored by the Office of Naval Research through the
Naval Research Laboratory 6.1 Computational Physics Task Area. 

\bibliographystyle{elsarticle-num}
\bibliography{../JCP_submission/citations}

\appendix

\section{Thermodynamic relations}

\label{sec:thermo-relations}

This section provides additional information on the thermodynamic
relations for a generic cubic equation of state written as
\begin{equation}
P=\frac{\widehat{R}T}{\widehat{v}-b}-\frac{a\alpha}{\left(\widehat{v}+\delta_{1}b\right)\left(\widehat{v}+\delta_{2}b\right)},\label{eq:generic-cubic-eos}
\end{equation}
where $\delta_{1}$ and $\delta_{2}$ are constants that depend on
the specific equation of state. Further details can be found in~\cite{Ma17,Ma18}.
For mixtures, the attractive and repulsive parameters (also known
as the energy and co-volume parameters, respectively) are computed
as
\begin{align*}
a\alpha & =\sum_{i=1}^{n_{s}}\sum_{j=1}^{n_{s}}X_{i}X_{j}a_{ij}\alpha_{ij},\\
b & =\sum_{i=1}^{n_{s}}X_{i}b_{i},
\end{align*}
and the critical mixture conditions for temperature, pressure, molar
volume, and acentric factor are given by
\begin{align*}
T_{c,ij} & =\sqrt{T_{c,i}T_{c,j}}\left(1-k_{ij}\right),\\
P_{c,ij} & =Z_{c,ij}\frac{\widehat{R}T_{c,ij}}{\widehat{v}_{c,ij}},\\
\widehat{v}_{c,ij} & =\frac{1}{8}\left(\widehat{v}_{c,i}^{1/3}+\widehat{v}_{c,j}^{1/3}\right)^{3},\\
\omega_{ij} & =\frac{\omega_{i}+\omega_{j}}{2},
\end{align*}
respectively, with $\left(\cdot\right)_{i}$ denoting the corresponding
property of the $i$th species. $k_{ij}$ is the binary interaction
parameter, which is assumed to be zero~\cite{Ma18}.

For the Peng-Robinson equation of state, $\delta_{1}=1+\sqrt{2}$,
$\delta_{2}=1-\sqrt{2}$, and
\begin{align*}
a_{ij} & =0.45724\frac{\widehat{R}^{2}T_{c,ij}^{2}}{P_{c,ij}},\\
\alpha_{ij} & =\left[1+c_{ij}\left(1-\sqrt{\frac{T}{T_{c,ij}}}\right)\right]^{2},\\
b_{i} & =0.07780\frac{\widehat{R}T_{c,ij}}{P_{c,ij}},\\
c_{ij} & =0.37464+1.54226\omega_{ij}-0.26992\omega_{ij}^{2}.
\end{align*}
In addition, we have 
\[
\frac{d\alpha_{ij}}{dT}=-c_{ij}\sqrt{\frac{\alpha_{ij}}{TT_{c,ij}}}
\]
and
\[
\frac{d^{2}\alpha_{ij}}{dT^{2}}=\frac{c_{ij}^{2}}{2TT_{c,ij}}+\frac{c_{ij}}{2}\sqrt{\frac{\alpha_{ij}}{T^{3}T_{c,ij}}}.
\]
Unless otherwise specified, the remainder of this section applies
to generic cubic equations of state of the form~(\ref{eq:generic-cubic-eos}).

Thermodynamic quantities are evaluated as the sum of the corresponding
mixture-averaged ideal-gas value (assuming a thermally perfect gas)
and a departure function~\cite{War95}. The ideal-gas values can
be computed based on polynomial fits~\cite{Mcb93,Mcb02} and are
denoted $\left(\cdot\right)^{(I)}$. The molar internal energy is
given by
\begin{align*}
\widehat{u}\left(T,\widehat{v},X_{i}\right) & =\widehat{u}^{(I)}\left(T,X_{i}\right)+\int_{\widehat{v}}^{\infty}\left[P-T\left(\frac{\partial P}{\partial T}\right)_{\widehat{v},X_{i}}\right]d\widehat{v}\\
 & =\widehat{u}^{(I)}\left(T,X_{i}\right)+K_{1}\left[a\alpha-T\left(\frac{\partial a\alpha}{\partial T}\right)_{X_{i}}\right],
\end{align*}
where
\[
K_{1}=\int_{\widehat{v}}^{\infty}\frac{d\widehat{v}}{\left(\widehat{v}+\delta_{1}b\right)\left(\widehat{v}+\delta_{2}b\right)}=\frac{1}{\left(\delta_{1}-\delta_{2}\right)b}\ln\left(\frac{\widehat{v}+\delta_{2}b}{\widehat{v}+\delta_{1}b}\right).
\]
The molar enthalpy can then be computed as
\[
\widehat{h}=\widehat{u}+P\widehat{v}.
\]
In this work, where we employ the Peng-Robinson equation of state
(i.e., $\delta_{1}=1+\sqrt{2}$ and $\delta_{2}=1-\sqrt{2}$), $K_{1}$
is calculated as
\[
K_{1}=\frac{1}{2\sqrt{2}b}\ln\left[\frac{\widehat{\mathsf{v}}+\left(1-\sqrt{2}\right)b}{\widehat{\mathsf{v}}+\left(1+\sqrt{2}\right)b}\right],\quad\widehat{\mathsf{v}}=\max\left\{ \widehat{v},b\right\} ,
\]
to prevent undefined behavior and for consistency with the constraint
$\widehat{v}>b$, which is motivated by the fact that for a pure fluid,
$b$ represents the molar volume as the pressure tends to infinity
(i.e., the co-volume)~\cite{Pin21}. In addition, the Helmholtz free
energy of a homogeneous mixture is undefined for $\widehat{v}\leq b$~\cite{Zha19}.
Enforcing this behavior directly on the solution in a rigorous manner
while maintaining accuracy is beyond the scope of this study.

The molar entropy is given by
\begin{align*}
\widehat{s}\left(T,\widehat{v},X_{i}\right) & =\widehat{s}^{(I)}\left(T,\widehat{v},X_{i}\right)+\int_{\widehat{v}}^{\infty}\left[\rho R-\left(\frac{\partial P}{\partial T}\right)_{\widehat{v},X_{i}}\right]d\widehat{v}\\
 & =\widehat{s}^{(I)}\left(T,\widehat{v},X_{i}\right)+\widehat{R}\ln\frac{\widehat{v}-b}{\widehat{v}}-K_{1}\left(\frac{\partial a\alpha}{\partial T}\right)_{X_{i}},
\end{align*}
where $R=\widehat{R}/\overline{W}$ is the specific gas constant,
with $\overline{W}$ denoting the mixture molecular mass, which is
defined as
\[
\overline{W}=\frac{\rho}{\sum_{i=1}^{n_{s}}C_{i}}=\sum_{i=1}^{n_{s}}X_{i}W_{i}.
\]
The molar heat capacities at constant volume and constant pressure
are evaluated as
\[
\widehat{c}_{v}=\left(\frac{\partial\widehat{u}}{\partial T}\right)_{\widehat{v},X_{i}}=\widehat{c}_{v}^{(I)}\left(T,X_{i}\right)-K_{1}T\left(\frac{\partial^{2}a\alpha}{\partial T^{2}}\right)_{X_{i}}
\]
and
\[
\widehat{c}_{p}=\left(\frac{\partial\widehat{h}}{\partial T}\right)_{P,X_{i}}=\widehat{c}_{p}^{(I)}\left(T,X_{i}\right)-\widehat{R}-K_{1}T\left(\frac{\partial^{2}a\alpha}{\partial T^{2}}\right)_{X_{i}}-T\frac{\left(\partial P/\partial T\right)_{\widehat{v},X_{i}}^{2}}{\left(\partial P/\partial\widehat{v}\right)_{T,X_{i}}^{2}},
\]
respectively. Finally, the square of the speed of sound is given by
\[
c^{2}=\left(\frac{\partial P}{\partial\rho}\right)_{s,X_{i}}=\frac{\gamma}{\rho\kappa_{T}},
\]
where $\gamma$ is the specific heat ratio and $\kappa_{T}$ is the
isothermal compressibility, written as
\[
\kappa_{T}=-\frac{1}{\widehat{v}}\left(\frac{\partial\widehat{v}}{\partial P}\right)_{T,X_{i}}.
\]
The partial derivatives in the above thermodynamic relations can be
expressed as follows:
\begin{align*}
 & \left(\frac{\partial a\alpha}{\partial T}\right)_{X_{i}}=\sum_{i=1}^{n_{s}}\sum_{j=1}^{n_{s}}X_{i}X_{j}a_{ij}\frac{d\alpha_{ij}}{dT},\\
 & \left(\frac{\partial^{2}a\alpha}{\partial T^{2}}\right)_{X_{i}}=\sum_{i=1}^{n_{s}}\sum_{j=1}^{n_{s}}X_{i}X_{j}a_{ij}\frac{d^{2}\alpha_{ij}}{dT^{2}},\\
 & \left(\frac{\partial P}{\partial T}\right)_{\widehat{v},X_{i}}=\frac{\widehat{R}}{\widehat{v}-b}-\frac{\left(\partial a\alpha/\partial T\right)_{X_{i}}}{\left(\widehat{v}+\delta_{1}b\right)\left(\widehat{v}+\delta_{2}b\right)},\\
 & \left(\frac{\partial P}{\partial\widehat{v}}\right)_{T,X_{i}}=-\frac{\widehat{R}T}{\left(\widehat{v}-b\right)^{2}}-\frac{a\alpha\left[2\widehat{v}+\left(\delta_{1}+\delta_{2}\right)b\right]}{\left(\widehat{v}+\delta_{1}b\right)^{2}\left(\widehat{v}+\delta_{2}b\right)^{2}}.
\end{align*}

\end{document}